\documentclass[fleqn,usenatbib]{mnras}

\usepackage{newtxtext,newtxmath}
\usepackage{caption}

\usepackage[T1]{fontenc}
\usepackage{ae,aecompl}
\makeatletter
\newcommand\footnoteref[1]{\protected@xdef\@thefnmark{\ref{#1}}\@footnotemark}
\makeatother
\usepackage{wrapfig,booktabs}
\usepackage{graphicx}	% Including figure files
\usepackage{amsmath}	% Advanced maths commands
\usepackage{amssymb}	% Extra maths symbols

\newcommand{\subsubsubsection}[1]{\paragraph{#1}\mbox{}\\}
\newcommand{\macc}{\textrm{GCU1}}
\newcommand{\AC}{\textrm{GCU2}}
\newcommand{\AD}{\textrm{GCU3}}
\newcommand{\AEp}{\textrm{GCU4}}

\newcommand{\RS}{\textrm{GCU5}}

\newcommand{\CC}{\textrm{GCU6}}

\newcommand{\irwin}{\textrm{GCU7}}

\newcommand{\BC}{\textrm{GCU8}}
\title[Globular Cluster ULXs]{X-Ray Spectral Variability of Ultraluminous X-Ray Sources in  Extragalactic Globular Clusters}

\author[K. C. Dage et al]{
Kristen C. Dage,$^{1}$\thanks{E-mail: kcdage@msu.edu}
Stephen E. Zepf,$^{1}$ 
Mark B. Peacock,$^{1}$ 
Arash Bahramian, $^{1, 2}$
\newauthor
Omid Noroozi, $^{1}$
Arunav Kundu, $^3$
Thomas J. Maccarone $^4$
\\
$^{1}$Department  of  Physics  and  Astronomy,  Michigan  State  University,  East
Lansing, MI 48824\\
$^{2}$ International Centre for Radio Astronomy Research $--$ Curtin University, GPO Box U1987, Perth, WA 6845, Australia \\
$^{3}$Eureka Scientific, Inc., 2452 Delmer Street, Suite 100 Oakland, CA 94602, USA\\
$^{4}$Department of Physics and Astronomy, Box 41051, Science Building, Texas Tech University, Lubbock, TX 79409-1051, USA
}

\date{Accepted XXX. Received YYY; in original form ZZZ}

\pubyear{2018}

\begin{document}

\label{firstpage}
\pagerange{\pageref{firstpage}--\pageref{lastpage}}
                       \maketitle

\begin{abstract}
A number of ultraluminous X-ray sources (ULXs) are physically associated with extragalactic globular clusters (GCs). We undertake a systematic X-ray analysis of eight of the brightest of these sources. %and compare these to the recently published analysis of a ninth such source.
We fit the spectra of the GC ULXs to single power law and single disk models. We find that the data never require that any of the sources change between a disk and a power law across successive observations. The GC ULXs best fit by a single disk show a bimodal distribution: they either have temperatures well below 0.5 keV, or variable temperatures ranging above 0.5 keV up to 2~keV. The GC ULXs with low kT have significant changes in luminosity but show little or no change in kT. By contrast, the sources with higher kT either change in both kT and $L_X$ together, or show no significant change in either parameter. Notably, the X-ray characteristics may be related to the optical properties of these ULXs, with the two lowest kT sources showing optical emission lines.
% * <mpeacock@msu.edu> 2018-07-03T16:25:51.884Z:
% 
% > The sources that have low inner disk temperatures  do not seem to vary much or at all with significant variations in luminosity.
% Maybe: The sources that have low inner disk temperatures exhibit little or no change in kT with significant variations in luminosity. 
% 
% ^ <kcdage@msu.edu> 2018-07-05T11:52:33.631Z.
% * <mpeacock@msu.edu> 2018-07-03T16:24:24.023Z:
% 
% > , which is present in the two low kT sources.
% maybe: with the two lowest kT sources showing optical emission lines (or whatever they show)
% 
% ^ <kcdage@msu.edu> 2018-07-05T11:52:32.250Z.
% * <mpeacock@msu.edu> 2018-07-03T16:16:07.864Z:
% 
% > The sources best fit by a single disk model either have temperatures well below 1 keV, or temperatures above 1 keV
% i know it says well below, but this sounds a little odd to me to say temperatures are either above 1kev or below (i.e. it has a temperature :) Maybe there's another way of phrasing this, not sure of the best way to edit. 
% 
% ^.

\end{abstract}


%%%%%%%%%%%%%%%%%%%%%%%%%%%%%%%%%%%%%%%%%%%%%%%%%%

%%%%%%%%%%%%%%%%% BODY OF PAPER %%%%%%%%%%%%%%%%%%

\section{INTRODUCTION}
\label{introduction}

Ultraluminous X-ray sources (ULXs) are non-nuclear X-ray sources with luminosities 
significantly greater than the Eddington limit or a 10  $M_\odot$ black hole (BH). $L_X \gtrsim 10^{39}$ erg s$^{-1}$ is often adopted as a general guideline for being well above this limit and therefore a ULX. This implies the accreting object in a ULX is either a black hole, or a neutron star (NS) with super-Eddington accretion and/or emission beamed along our line of sight. The most well studied ULX population is that in star forming galaxies (e.g. review by \citealt{2017ARA&A..55..303K}). The ULXs with luminosities lower than 3 $\times 10^{39}$ erg s$^{-1}$  are typically fit best by a singly peaked, broadened disk.  The higher luminosity ULXs are fit by a two component disk and power law model, and are either ``soft"  or ``hard", depending on the slope of the power law. The slope differentiates between $\sim$ Eddington and super-Eddington models  \citep{2009MNRAS.397.1836G, 2013MNRAS.435.1758S}.  These ULXs show strong luminosity variability that are often accompanied by significant changes in states/spectral shapes \citep{ 2013MNRAS.435.1758S,2013ApJ...778..163B, 2013ApJ...779..148W, 2016ApJ...831..117F}. A third group of ULXs have also been identified, with luminosities below 3 $\times 10^{39}$ erg s$^{-1}$  and soft blackbody dominated emission and are known as  of ``supersoft'' ULXs \citep{2016MNRAS.456.1859U}.

Some optical counterparts  have been identified for ULXs in star forming regions. These are primarily OB type giants or supergiants \citep[][and references therein]{2018MNRAS.477L..90P}.  High mass companion stars could provide an explanation for the high mass transfer rates which drive the high luminosities and account for their connection to star-forming regions (see \citealt{2009MNRAS.397.1836G} and references therein).

Of the many sources confirmed as ULXs, a number of those associated with star forming galaxies have been found to have pulsations, implying that the compact object is a neutron star (e.g., \citealt{2014Natur.514..202B} to  \citealt{2018MNRAS.477L..90P}).  Although ULXs in star forming galaxies have been studied extensively,  ULXs also exist in a completely different environment: globular clusters (GCs).  Since 2007, five globular cluster ULX sources have been studied in some depth: XMMUJ122939.7+075333 \citep[RZ2109;][]{2007Natur.445..183M}, CXOJ0338318-352604 \citep{irwin2010}, CXOKMZJ033831.7$-$353058 \citep{2010ApJ...721..323S}, CXOU 1229410+0757442 \citep{2011MNRAS.410.1655M}, and CXOUJ1243469+113234 \citep{2012ApJ...760..135R}.  Some of these sources are proposed to have black hole primaries due to their highly variable ULX emission and other properties
\citep{2007Natur.445..183M,2010ApJ...721..323S}.

%\textbf{\citet{2016Natur.538..356I} also identified two globular clusters (near NGC 4697) that were associated with ultraluminous X-ray flares. The flares were brief and decayed over the course of an hour. By contrast, the five sources mentioned above stayed at high X-ray luminosities over the course of many \textit{Chandra} observations. }

% 
Interestingly, nebular emission is observed from some of these GC ULXs. Such emission is likely associated with the ULX, since it is extremely rare in globular clusters, which lack young stars and are known to have few planetary nebulae \citep[and references therein]{2012ApJ...759..126P}. Their forbidden optical emission lines limit beaming to a factor of a few or less for some of these GC sources.  While geometric beaming has been proposed as a mechanism for producing the super-Eddington X-ray emission from accreting neutron stars \citep{2001ApJ...552L.109K},  \citet{peacock2012a} find that the optical emission from the GC ULX sources are too luminous to accommodate a large beaming factor and should be isotropically emitting. (See also \citet{2002astro.ph..2488P}, and \citet{binder18} for a discussion of beaming in ULX sources.)

The presence of black holes in globular clusters, and if so, the properties the globular cluster black holes has been debated for many decades (see \citealt{spitz1969}, \citealt{chatterjee17}, \citealt{2017MNRAS.469.4665P}). While black holes are initially expected to form in globular clusters \citep{ivanova2010}, early numerical simulations indicated that dynamical interactions within the globular cluster would cause the black holes to be ejected \citep{kulkarni1993,1993Natur.364..423S}. However, more recent work, (e.g., \citealt{2013ApJ...763L..15M, 2015ApJ...800....9M,heggie14,2013MNRAS.430L..30S}) show that the ejection time scales are much longer, and that stellar mass black holes remain well mixed in the cluster.  In addition to this, globular clusters are prime environments for black hole mergers, and if black holes are retained in globular clusters, then globular clusters could be the  progenitors of recent LIGO detections of merging black holes \citep[see][]{abbottb,2016PhRvD..93h4029R}. Studying GC ULX sources could shed some light on the nature of black holes in globular clusters. 

% * <mpeacock@msu.edu> 2018-07-03T17:05:34.049Z:
% 
% changed this a fair bit - think the old version is saved if you dont like it
% 
% ^ <kcdage@msu.edu> 2018-07-05T11:53:33.117Z.
It is interesting to compare the ULX population in star forming regions to those in globular clusters, which are two very different environments. Firstly, the dense globular cluster environment is conducive to dynamical formation of binary systems, thus, GC ULXs are likely dynamically formed \citep{ivanova2010}. The large interaction cross-section in this environment means that it is unlikely that the current binary partner of any compact object in a GC was a binary with the object when the stars initially formed and evolved. In contrast, ULX sources in star forming regions likely evolved from primordial binaries, and that the two objects in the binary evolved at the same time.

The donor stars in the two types of ULXs are also likely to be very different; due to the ages of globular clusters, 13 Gyr or so,  the BHs and NSs were born from the massive stars in these GCs many Gyr ago. This very old age for these primordial BH and NS populations differs greatly from the young ages of the NSs and BHs formed from massive stars in currently star forming regions. For example, one such GC ULX source likely has a white dwarf as its donor star \citep{2014ApJ...785..147S}, but the ULXs in star-forming regions should have high mass donor stars, as mentioned previously. 

One other difference between these ULX populations is that those in star forming regions have hydrogen emission present in their optical spectra \citep[e.g.,][]{2015NatPh..11..551F}, while at least three of the globular cluster ULXs have no hydrogen emission \citep{zepf08,irwin2010, 2012ApJ...760..135R}. The most well studied GC ULX, RZ2109, also has very different X-ray behaviour from ULXs in star forming regions; it varies by more than an order of magnitude in X-ray luminosity between many different observations, but exhibits little or no variation in kT over those same observations  \citep{2008MNRAS.386.2075S, 2018arXiv180601848D}. 

Several studies have considered the nature of the GC ULX sources and their optical emission, although most of them tend to focus on the source RZ2109 \citep[e.g.,][]{zepf08,steele11, peacock2012a,2012ApJ...759..126P,  2014ApJ...785..147S}. 
There are also studies of the optical spectrum of 
the GC ULX CXOJ0338318-352604 in the galaxy NGC~1399 \citep{irwin2010}. The emission lines of RZ2109 have been variously modelled as a $\sim 50 - 100 $ $M_\odot$  mass black hole tidally disrupting a horizontal branch star \citep{clausen12} and as the ejecta from a R Corona Borealis star being photoionised by an unrelated X-ray source elsewhere in the cluster \citet{2011MNRAS.410L..32M}. The goal of this paper is to consider all known GC ULX sources to better understand their nature and constrain these models.

%SEZ: I rewrote above. Old version is here. Another GC ULX in NGC~1399 \citep{irwin2010} has been modeled by \citet{clausen12} as an intermediate mass black hole tidally disrupting a horizontal branch star. \citet{2011MNRAS.410L..32M} model the same system as ejecta from an R Corona Borealis star being photoionized by an X-ray source elsewhere in the cluster. The goal of this paper is to consider all known GC ULX sources to better understand their nature and constrain these models.

In this paper, we undertake an analysis of the eight known globular cluster ULXs with L$_X > 10^{39}$ erg s$^{-1}$ to broaden our understanding of these sources. As noted above, three of these sources have been studied previously, while five are new to this paper. Section \ref{sec:observations} discusses the \textit{Chandra} observations and analysis, the results are presented in Section \ref{sec:results}. The major results of the paper are discussed further in Section \ref{sec:conclusions}.

\section{Data and Analysis}
\label{sec:observations}
\subsection{Globular Cluster ULX Sample}

We consider low mass X-ray binaries (LMXBs) in the sample of seven local early-type galaxies as presented in \citet{Peacock14}.  The galaxies studied are all within 20~Mpc and have deep Chandra observations of $>100$~ksec. We select the targets that are located in GCs in these galaxies and that have X-ray luminosities greater than $10^{39}$ erg s$^{-1}$. The X-ray fluxes of the point sources in these galaxies are published by \citealt{Paolillo11} (NGC~1399), \citealt{Brassington08} (NGC~3379),  \citealt{Brassington09} (NGC~4278),  \citealt{Joseph13} (NGC~4472), \citealt{Li10} (NGC~4594), \citealt{Luo13} (NGC~4649), and \citealt{Sivakoff08} (NGC~4697). Globular cluster X-ray sources were then identified from these catalogues by matching to optical counterparts in aligned HST optical images \citep[see][for details]{Peacock14, Luo13}. The resulting catalogue of high probability GC LMXB candidates is used to select ULX sources. 

%We consider \textbf{all of the currently known} globular cluster ULX sources with average $L_X \gtrsim 10^{39}$ erg s$^{-1}$. 
%Four of these samples have been previously studied in depth: NGC 4472 \citep{2007Natur.445..183M,2011MNRAS.410.1655M}, NGC 4949 \citep{2012ApJ...760..135R}, and NGC 1399 \citep{irwin2010}. 
\citet{2007ApJ...660.1246S} predict that the X-ray luminosity of bright GC sources is dominated by a single high luminosity object and not from multiple fainter sources with a combined high luminosity.  Many of the sources in our sample are also highly variable in X-ray, which also implies that most or all of  the luminosity comes from a single source (see \citealt{2007Natur.445..183M} and Section \ref{appendix-var}  for further discussion of source variability). See Table \ref{all_srcs} for source coordinates and optical cluster properties. We present new data and analysis of three of the previously studied sources (\citealt{irwin2010,2011MNRAS.410.1655M,2012ApJ...760..135R}), and new analysis of five GC ULXs which have not yet been previously studied in depth.

\subsection{Spectral Fitting}
 We use archival \textit{Chandra} data of NGC~1399, NGC~4472 and NGC~4649 (See Tables \ref{obs4472}. For these observations, we use \textsc{ciao-4.9}\footnote{\url{http://cxc.harvard.edu/ciao/}}'s  \citep{2006SPIE.6270E..1VF} \texttt{specextract} function to extract the spectra, with approximately 2.5$^{\prime\prime}$ circular regions on the sources, and set a series of 5-10 similarly shaped regions in source-less areas around the sources to select the background regions.  We follow \citet{2018arXiv180601848D} for all fitting of \textit{Chandra} spectra. Observations with counts greater than 100 were binned by 20, those with less were binned by 1. We fit the spectra with  with \textsc{xspec}\footnote{\url{https://heasarc.gsfc.nasa.gov/xanadu/xspec/}} \citep{1996ASPC..101...17A}, using $\chi^2$ statistics for the more detailed spectra and C-statistics\footnote{\url{https://heasarc.gsfc.nasa.gov/xanadu/xspec/manual/XSappendixStatistics.html}} \citep{1979ApJ...228..939C} for the spectra which were binned in counts of 1. We set the abundance of elements to Wilms \citep{2000ApJ...542..914W}, and freeze the value of the equivalent hydrogen column absorption (nH) to the value for that galaxy\footnote{\url{http://cxc.harvard.edu/toolkit/colden.jsp}}. We use the ``ignore bad" command to remove bad channels.

All of the data were fit with two separate single component models. The first is a multi-temperature blackbody disk (\texttt{tbabs*diskbb}) \citep{1984PASJ...36..741M}. The second is a pegged power law model (\texttt{tbabs*pegpwrlw}) with the normalisation pegged from 0.5-8.0~keV. We also fit the high count (>100 counts) data with a two component model \texttt{tbabs*(diskbb+pegpwrlw)}, and used F-test to determine if any improvement was statistically significant. Lastly, we determine to what extent (if any) there is intrinsic absorption in these systems by fitting a second absorbing column to the high count data (\texttt{tbabs*tbabs*pegpwrlw} and \texttt{tbabs*tbabs*diskbb}.)

The \texttt{pegpwrlw} model is normalised such that the best fit to the power law norm is the unabsorbed flux from 0.5-8.0 keV. To determine the unabsorbed flux in the models fit by \texttt{diskbb},we multiply the models by \texttt{cflux}, with the energy range between 0.5 and 8 keV and fit. To calculate the luminosities, we use the distances of 20.0 Mpc for NGC 1399 \citep{2001MNRAS.327.1004B}, 16.8 Mpc for NGC 4472 \citep{macri} and 16.5 Mpc for NGC 4649 \citep{2009ApJ...694..556B}.
% * <mpeacock@msu.edu> 2018-07-03T17:38:50.272Z:
% 
% > All of the data were fit with two separate single component models, \texttt{tbabs*diskbb} and \texttt{tbabs*pegpwrlw} (pegged from 0.5-8keV).
% I'd describe these models, not just the code names ie: 
% All of the data were fit with two separate single component models, a power law (\texttt{tbabs*diskbb}) and a disk black body (\texttt{tbabs*pegpwrlw}, pegged from 0.5-8keV) ***if those are the models!
% 
% ^ <kcdage@msu.edu> 2018-07-05T11:56:06.024Z.

\begin{table*}
\begin{minipage}{0.75\textwidth}
\caption{Coordinates and optical properties of the GC ULX sample. Optical properties from \citep[][RZ2109]{2007ApJ...669L..69Z}, \citep[][GCU1]{MKZ03}, \citet[][GCU2, GCU3, GCU4]{Peacock14}, \citep[][GCU5, GCU6]{Strader12} and  \citep[][GCU7, GCU8]{Paolillo11} All magnitudes are in $z$-band unless otherwise noted. }
\label{all_srcs}
\centering
\begin{tabular}{llllll}
\hline
\hline
Object & RA      & Dec      & $z$    & $g-z$    & Host Galaxy (Distance)   \\ \hline
RZ2109 & 12:29:39.9 & +07:53:33.3 & 20.4 \footnote{\label{note1}Converted from V to $z$ by using the relation V = $g$ - 0.39($g-z$) + 0.07 \citep{2010MNRAS.407.2611P}.} & 0.84 \footnote{$g-z$ conversion from B-R used the following relationship: $g-z$ = 1.305(B-R) - 0.543 \citep{2010MNRAS.407.2611P}. } & NGC 4472 (16.8 Mpc)\footnote{\label{note2}Distance from \citealt{macri}}\\ 
GCU1   & 12:29:41.0 & +07:57:44.2   & 20.8 \footnoteref{note1}   & 1.59 \footnote{$g-z$ conversion from V-I used the following relationship: $g-z$ = 1.518(V-I) - 0.443 \citep{2010MNRAS.407.2611P}.} & NGC 4472 (16.8 Mpc)\footnoteref{note2}\\ 
GCU2   & 12:29:34.5 & +08:00:32.1   & 22.1              & 0.92      & NGC 4472 (16.8 Mpc)\footnoteref{note2}\\ 
GCU3   & 12:29:42.3 & +08:00:08.1    & 19.5              & 1.42     & NGC 4472 (16.8 Mpc)\footnoteref{note2}   \\ 
GCU4   & 12:29:34.5 & +07:58:51.6  & 20.1              & 1.11    & NGC 4472   (16.8 Mpc)\footnoteref{note2}  \\ 
GCU5   & 12:43:46.9 & +11:32:34   & 20.3 &1.55  & NGC 4649 (16.5 Mpc)\footnote{\label{note3}Distance from \citealt{2009ApJ...694..556B}} \\
GCU6   & 12:43:44.5 & +11:31:50     & 22.2             & 1.60       & NGC 4649 (16.5 Mpc)\footnoteref{note3} \\ 
GCU7   & 03:38:31.8 & -35:26:04    & 20.7               & 1.98     & NGC 1399 (20.0 Mpc)\footnote{\label{note4}Distance from \citealt{2001MNRAS.327.1004B}}    \\ 
GCU8   & 03:38:32.6 & -35:27:05.7 & 19.9               & 2.24     & NGC 1399   (20.0 Mpc)\footnoteref{note4} \\ \hline
\end{tabular}
\end{minipage}
\end{table*}

\subsubsection{NGC 4472 GC ULXs}
\begin{figure*}

\includegraphics[scale=0.45]{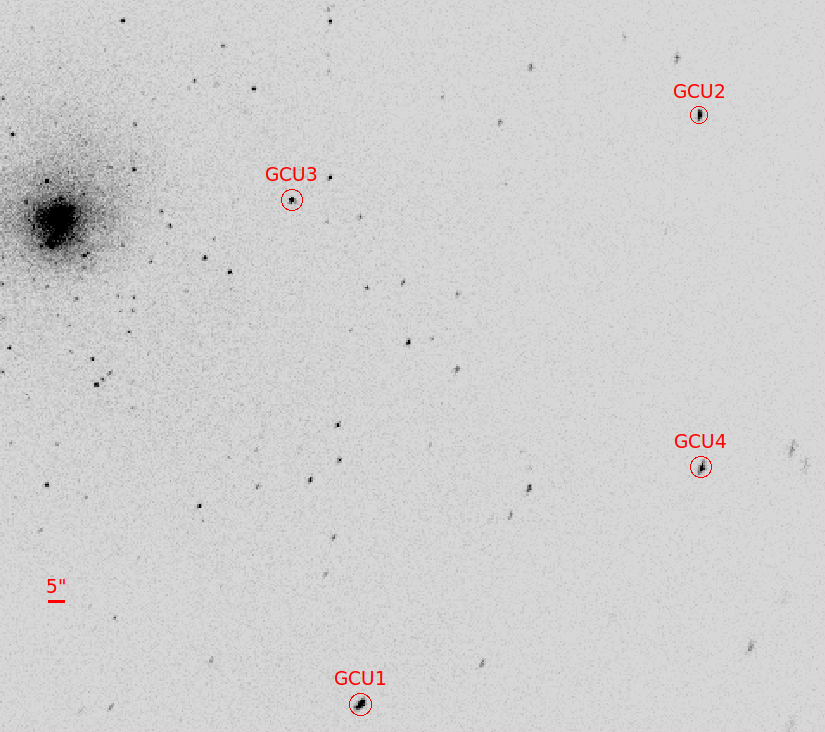}

\caption{X-ray image of NGC 4472 (ObsID 12888, filtered to 0.5-8.0 keV) with regions for GCU1-GCU4 overlaid. } 
\label{fig:img4472}
\end{figure*}

There are five globular cluster ULX sources in NGC~4472 with $L_X > 10^{39}$ erg s$^{-1}$. The brightest of these, RZ2109 has been previously well studied in both X-ray and optical (\citealt{2007Natur.445..183M}, \citealt{zepf08}, \citealt{2008MNRAS.386.2075S}, \citealt{2011ApJ...739...95S}, \citealt{2018arXiv180601848D}). A second globular cluster ULX, CXOU 1229410+0757442, hereafter $\macc$, has also been studied in some detail by \citet{2011MNRAS.410.1655M}. There are three other GC ULXs, CXOU 1229345+08003209, (hereafter $\AC$), CXOU 1229423+08000808 (hereafter $\AD$), CXOU1229345+07585155 (hereafter $\AEp$), which have not yet been studied in depth (see Figure \ref{fig:img4472} for a \textit{Chandra} image of the source locations.). At least 10 different \textit{Chandra} observations exist for these sources spanning from the year 2000 to 2016 (see Table \ref{obs4472}), enabling a study of their spectral properties and behaviour  on the scale of years to decades. Below, we discuss the spectral fitting results for all sources. 

\begin{table}
\centering
\caption{Observations of NGC 4472, with raw source counts (0.5-8.0 keV)  for $\macc$, $\AC$, $\AD$  and $\AEp$. Observations marked with * were too off axis to measure counts from. Observations marked with - had the source off the chip.}
\label{obs4472}
\begin{tabular}{lclclclclclclclc}
\hline
\hline
ObsID & Date       & ObsLen & GCU1 & GCU2 & GCU3 & GCU4  \\ 
&&(ks) &  Cts &  Cts &  Cts &  Cts \\\hline
322   & 2000-03-19 & 10.36   &42&34&15&  41  \\ 
321   & 2000-06-12 & 39.59    &134&142&178& 196  \\ 
8095  & 2008-02-23 & 5.09     &31&11&11&21   \\ 
11274 & 2010-02-27 & 39.67    &458&115&196& 180 \\ 
12978 & 2010-11-20 & 19.78    &12&- &2&*  \\ 
12889 & 2011-02-14 & 135.59    &1067&492&491&488  \\ 
12888 & 2011-02-21 & 159.31   &1559&506&644& 641  \\ 
16260 & 2014-08-04 & 24.74     &6&*&*&50  \\
16261 & 2015-02-24 & 22.76     &58&42&*& * \\ 
16262 & 2016-04-30 & 24.73      &136&50&89&52 \\ \hline
\end{tabular}
\end{table}

\subsubsubsection{CXOU 1229410+0757442 (GCU1)}

For $\macc$,  \citet{2011MNRAS.410.1655M} have previously examined earlier data through ObsID 11274, and could not find a statistical difference between an absorbed power law model versus an absorbed disk model, preferring the disk model for physical reasons.  While the lower count spectra were ambiguous as to whether a single disk model is a better fit than a single power law model,   $\chi^2$ statistics for the deep observations (ObsIDs 12888 \& 12889) indicate that the disk model alone  is a much better fit than a single power law. Values for both models are presented in Table \ref{4472fits}.% (Table \ref{4472B}). 

We fit a power law with a second absorbing column (\texttt{tbabs*tbabs*pegpwrlw}) to observations with greater than 100 source counts. We find that the nH is not inconsistent with zero in three of these observations.  By comparison, fitting \texttt{tbabs*tbabs*diskbb} resulted in the second absorbing column consistent with zero that had significantly better $\chi^2$ values in all cases.

We also find that adding a power law component to the single component disk is not a statistically significant improvement to the fit of $\macc$. Specifically, we used \textsc{xspec}'s F-test tool \footnote{https://heasarc.gsfc.nasa.gov/xanadu/xspec/manual/node83.html} which found no statistically significant improvement in the fit from adding the second component.  %\textsc{xspec}'s F-test tool \footnote{https://heasarc.gsfc.nasa.gov/xanadu/xspec/manual/node83.html} showed that any  improvement to the fit due to the second component is very likely to be due to random fluctuations.
%Conversely, if we first fit these data using a power law, adding a MCD component is generally shown to significantly improve the fit statistics.
We compare the statistics of the two component model to both single component models; these values are presented in Table \ref{4472_ftest}.  We note that while ObsID 11274 seems to favor a two component model, the best fit two-component power law photon index is 6.8, which is not typical for X-ray binaries. We do not report the best fit values for the two-component models as they are not physically  realistic or statistically significant. %\citet{MKZ03} found the optical properties of the cluster to be: V magnitude = 21.77 and V - I = 1.34.

%\begin{table*}
%\centering
%\caption{F-test probability values for single component versus two component models for observations of $\macc$ with over 100 source counts. We compare statistics between \texttt{tbabs*(diskbb+pegpwrlw)} and \texttt{tbabs*diskbb} only in columns titled "Disk", and \texttt{tbabs*(diskbb+pegpwrlw)}  with \texttt{tbabs*pegpwrlw} only in columns titled "PL".  }
%%\label{4472_ftest}
%\begin{tabular}{l|ll|ll}
%\hline
%\hline
%ObsID & Disk  & PL   \\\hline
%321   & 1      & 0.31               \\  
%11274 & 0.07       & 0.06                    \\
%12889 & 0.18      & 5.65e-8                            \\ 
%12888 & 0.14    & 1.37e-7       \\ \hline 
%\end{tabular}
%\end{table*}

\subsubsubsection{CXOU 1229345+08003209 (GCU2)}
$\AC$ had consistently better fit statistics for a  single component disk model (Table \ref{4472fits}).  The F-test values (Table \ref{4472_ftest}) again indicate that a single component disk model is the best fit for this source. 

We rule out the necessity of a second absorbing column for $\AC$ by again fitting \texttt{tbabs*tbabs*pegpwrlw} and \texttt{tbabs*tbabs*diskbb} to the source. Only ObsIDs 11274 and 12888 have a second absorption component for the power law model that is non-zero, however, the disk model,which has the best fit second absorbing column consistent with zero, has significantly better fit statistics than the intrinsically absorbed power law.

%\begin{table*}
%\centering
%\caption{F-test probability values for single component versus two component models for observations of $\AC$ with over 100 source counts. We compare statistics between \texttt{tbabs*(diskbb+pegpwrlw)} and \texttt{tbabs*diskbb} only in columns titled "Disk", and \texttt{tbabs*(diskbb+pegpwrlw)}  with \texttt{tbabs*pegpwrlw} only in columns titled "PL".}
%\label{4472_ftest2}
%\begin{tabular}{l|ll|ll|ll|ll|ll}
%\hline
%\hline
%ObsID & Disk  &  PL    \\\hline
%321        & 1& 0.06                         \\  
%12889 &0.10 & 0.17       \\ 
%12888   &0.36 & 5.83e-4                   \\ \hline 
%\end{tabular}
%\end{table*}
\subsubsubsection{CXOU 1229423+08000808 (GCU3)}

We fit a power law model to longer observations of $\AD$ with a second absorbing column, but found the best fit absorption to be consistent with zero in all cases.
We fit $\AD$ with both single component models (as reported in Table \ref{4472fits}), as well as a two component model. The statistics generally favour a single component disk fit, or are inconclusive. The two component model only showed a significant statistical improvement in one observation (Table \ref{4472_ftest}), however, the best fit two-component power law for that had a very high index of 5.7, which again is unphysical and much steeper than the typical value for X-ray binaries.

%\begin{table*}
%\centering
%\caption{F-test probability values for single component versus two component models for observations of  $\AD$ with over 100 source counts. We compare statistics between \texttt{tbabs*(diskbb+pegpwrlw)} and \texttt{tbabs*diskbb} only in columns titled "Disk", and \texttt{tbabs*(diskbb+pegpwrlw)}  with \texttt{tbabs*pegpwrlw} only in columns titled "PL".}
%\label{4472_ftest3}
%\begin{tabular}{l|ll|ll|ll|ll|ll}
%\hline
%\hline
%ObsID & Disk  & PL   \\\hline
%321    & 0.01 & 0.99                \\  
%12889 & 2.31e-3 & 4.07e-3                           \\ 
%12888 &0.10& 6.01e-6     \\ \hline 
%\end{tabular}
%\end{table*}
\subsubsubsection{CXOU1229345+07585155 (GCU4)}
We found that the best fit value for the second absorbing column of this source was consistent with zero across all of the longer observations. $\AEp$ was statistically better fit by \texttt{tbabs*pegpwrlw} (Table \ref{4472fits}). The F-test values generally do not favour a two component model (Table \ref{4472_ftest}), except for ObsID 12889. However, it again has an unphysical power law of 5.0.

\begin{table*}
\caption{\textit{Chandra} Fit Parameters and Fluxes (0.5-8 keV) for spectral best fit single-component models, \texttt{tbabs*diskbb} and \texttt{tbabs*pegpwrlw} for GC ULXs in NGC 4472. Hydrogen column density ($N_H$) frozen to 1.6 $\times 10^{20}$  cm$^{-2}$). Fit parameters marked with $^{**}$ either encountered an \textsc{xspec} error when computing a lower bound, or had a lower bound consistent with zero, and are presented as an upper limit. Lower count observations fit with C-stat have their statistics presented in parentheses. All fluxes shown are unabsorbed. }
\label{4472fits}
\begin{tabular}{|l|clc|lclc|}
\hline
\hline
\toprule
  &    & & $\macc$ \\ \hline
  &       \multicolumn{3}{c}{\textsc{tbabs*pegpwrlw}} & \multicolumn{3}{|c}{\textsc{tbabs*diskbb}} \\ \hline
\midrule
 ObsID/Date  & $\Gamma$ &   $\chi^2_{\nu}$/d.o.f. & PL Flux & $T_{in}$ &  Disk Norm &  $\chi^2_{\nu}$/d.o.f. &  Disk Flux  \\
&& \textbf{or} (C-stat)&($10^{-14}$ erg cm$^{-2}$ s $^{-1})$&(keV)&($10^{-4}$)& \textbf{or} (C-stat)&($10^{-14}$ erg cm$^{-2}$ s $^{-1})$\\ \hline  \hline

 322 (2000-03-19) &1.7 ($\pm$ 0.5)& (22.74/36)&4.5 ($^{+2.5}_{-1.5})$&  0.75 ($^{+0.33}_{-0.25})$ & $\leq$ 136.13$^{**}$ &(26.83/36)&2.9 ($^{+1.1}_{-0.8})$ \\ 
321 (2000-06-12) &1.7 ($\pm$ 0.2)&3.41/5&2.6 ($\pm$ 0.6)&0.98   ($^{+0.24}_{-0.20})$ &12.9 ($^{+15.9}_{-7.2})$& 1.51/5 & 2.3 ($\pm 0.5)$\\
%2008-02-23 & ---&---&N/A& $\leq$1.61 $\times 10^{-14}$  \\
 11274 (2010-02-27) &1.3 ($\pm$ 0.1)&1.77/22&12.8 ($\pm$ 1.4)&1.7 ($^{+0.33}_{-0.25})$   &7.7 ($^{+5.5}_{-3.4})$&1.74/22&  11.1 ($\pm 1.5$) \\ 
 %2010-11-20 &---\footnote{fit with fixed \textbf{power law} of 1.6} &---&  N/A&  2.61 ($^{+17.02}_{-12.99})$ $\times$ $10^{-15}$  \\ 
12889 (2011-02-14)  &1.6 ($\pm$ 0.10)& 2.25/49& 6.8 ($\pm$ 0.5)&1.13 ($\pm 0.10$) &18.5 ($^{+6.5}_{-4.8})$& 1.25/49 &5.8($\pm 0.5$)\\ 
12888 (2011-02-21) &1.5 ($\pm$ 0.1)&1.71/69&8.8 ($\pm$ 0.5) &1.29 ($\pm 0.1$) & 14.1 ($^{+4.1}_{-3.2})$ & 1.10/69& 7.6($\pm 0.5)$\\ 
 %2014-08-04   &---\footnote{fit with fixed \textbf{power law} of 1.6}   &--- & N/A & 1.84($^{+13.20}_{-9.56})$$\times$$10^{-15}$\\ 
 16261 (2015-02-24) &2.0 ($\pm$ 0.4)&(47.61/51)&3.3  ($^{+0.9}_{-0.7}$)& 0.87  ($^{+0.29}_{-0.19}$)& $\leq$ 64.01$^{**}$& (41.04/51)  & 2.7($^{+0.7}_{-0.6})$\\ 
 16262 (2016-04-30)  &1.8 ($\pm 0.3$)&0.73/5&5.9 ($\pm$ 1.0)& 0.98  ($^{+0.29}_{-0.22})$ &26.9($^{+34.0}_{-15.2})$&   0.55/5 &4.9($\pm 0.9$)\\ 

\toprule
  &    & & $\AC$ \\
\midrule
 ObsID/Date  & $\Gamma$ &   $\chi^2_{\nu}$/d.o.f.  & PL Flux  & $T_{in}$ &  Disk Norm &  $\chi^2_{\nu}$/d.o.f.  &  Disk Flux \\
&& \textbf{or} (C-stat)&($10^{-14}$ erg cm$^{-2}$ s $^{-1})$&(keV)&($10^{-4}$)& \textbf{or} (C-stat)&($10^{-14}$ erg cm$^{-2}$ s $^{-1})$\\ \hline  \hline

 322 (2000-03-19) &1.4 ($\pm$0.5)&(35.31/32)& 3.9 ($^{+1.9}_{-1.3})$&  1.29 ($^{+1.01}_{-0.42})$ & $\leq$ 24.2 $^{**}$ &(31.83/32)& 3.2 ($^{+1.6}_{-1.1})$ \\ 
321 (2000-06-12) &1.3 ($\pm$ 0.2)&3.11/6&3.0 ($\pm$ 0.7)& 1.36   ($^{+0.39}_{-0.29})$ &  3.8 ($^{+4.4}_{-2.1})$& 0.70/6 & 2.5 ($\pm 0.6)$  \\
%2008-02-23 & ---&---&N/A& $\leq$1.61 $\times 10^{-14}$  \\
 11274 (2010-02-27) &1.4 ($\pm$ 0.3)&3.27/5&2.9 ($\pm$ 0.7)&1.45 ($^{+0.56}_{-0.37})$   & $\leq$ 7.89$^{**}$&1.35/5& 2.5 ($\pm 0.7$)  \\ 
 %2010-11-20 &---\footnote{fit with fixed \textbf{power law} of 1.6} &---&  N/A&  2.61 ($^{+17.02}_{-12.99})$ $\times$ $10^{-15}$  \\ 
12889 (2011-02-14)&1.4 ($\pm$ 0.1)&9.29/24&3.5 ($\pm$ 0.4)&1.37 ($\pm 0.20$) & 4.5($^{+2.7}_{-1.7})$& 1.11/24 &3.0($\pm 0.4$ )\\ 
12888 (2011-02-21) &1.6 ($\pm$ 0.1)&965459.3/24& 3.2 ($\pm$ 0.3) & 1.16 ($\pm 0.17$) &7.3 ($^{+4.7}_{-2.9})$ & 1.36/24& 2.6($\pm 0.3)$\\ 
 %2014-08-04   &---\footnote{fit with fixed \textbf{power law} of 1.6}   &--- & N/A & 1.84($^{+13.20}_{-9.56})$$\times$$10^{-15}$\\ 
 16261 (2015-02-24) &1.6 ($\pm$ 0.5)&(38.69/37)&2.5 ($^{+0.9}_{-0.7}$)& 1.25 ($^{+0.78}_{-0.36}$)&$\leq$ 5.67$^{**}$ & (36.00/37)& 2.1($^{+0.8}_{-0.6})$\\ 
  16262 (2016-04-30)  &1.9 ($\pm$ 0.5)&(36.02/49)&2.91 ($^{+0.91}_{-0.72}$)& 1.10  ($^{+0.63}_{-0.32}$) &$\leq$ 32.0$^{**}$&  (37.53/49)&2.40 ($^{+0.87}_{-0.63}$)\\ 

\hline

\toprule
  &  &  & $\AD$ \\ \hline
\midrule
 ObsID/Date  & $\Gamma$ &   $\chi^2_{\nu}$/d.o.f.  & PL Flux  & $T_{in}$ &  Disk Norm &  $\chi^2_{\nu}$/d.o.f.  &  Disk Flux  \\
&& \textbf{or} (C-stat)&($10^{-14}$ erg cm$^{-2}$ s $^{-1})$&(keV)&($10^{-4}$)& \textbf{or} (C-stat)&($10^{-14}$ erg cm$^{-2}$ s $^{-1})$\\ \hline  \hline

321 (2000-06-12) &1.3 ($\pm 0.2$)& 0.75/8&  3.7 ($\pm$ 0.7) & 1.41  ($^{+0.40}_{-0.30})$ &  4.1 ($^{+4.7}_{-2.3})$& 0.43/8 & 3.2 ($\pm 0.7)$ \\
%2008-02-23 & ---&---&N/A& $\leq$1.61 $\times 10^{-14}$  \\
 11274(2010-02-27)& 1.4 $(\pm$ 0.2) &2.56/8& 4.5 ($^{+0.8}_{-0.7})$  &1.19 ($^{+0.26}_{-0.20})$   &9.5($^{+8.3}_{-4.7})$&1.36/8& 3.6($\pm 0.6$) \\ 
 %2010-11-20 &---\footnote{fit with fixed \textbf{power law} of 1.6} &---&  N/A&  2.61 ($^{+17.02}_{-12.99})$ $\times$ $10^{-15}$  \\ 
12889 (2011-02-14) &1.4 ($\pm$0.1)& 0.90/23& 3.3 ($\pm$ 0.4) &1.39 ($^{+0.25}_{-0.20})$ &  3.9 ($^{+2.8}_{-1.7})$& 0.95/23 & 2.9 ($\pm 0.4)$  \\
12888 (2011-02-21) &1.5 ($\pm$1.0)&1.64/30& 3.6 ($\pm$ 0.3) & 1.31 ($\pm 0.15$) & 5.4($^{+2.6}_{-1.8})$& 0.92/30& 3.1($\pm 0.3)$\\ 
 %2014-08-04   &---\footnote{fit with fixed \textbf{power law} of 1.6}   &--- & N/A & 1.84($^{+13.20}_{-9.56})$$\times$$10^{-15}$\\ 
 16262 (2016-04-30) &1.5 ($\pm$ 0.4)&(81.63/77)&4.4 $(^{+1.3}_{-1.0})$& 1.27 ($^{+0.62}_{-0.32}$)& $\leq$ 20.9$^{**}$& (78.08/77)  & 3.6($^{+1.1}_{-0.8})$\\ 
\hline

\toprule
  &      & & $\AEp$ \\ \hline
\midrule
 ObsID/Date  & $\Gamma$ &   $\chi^2_{\nu}$/d.o.f.  & PL Flux  & $T_{in}$ &  Disk Norm &  $\chi^2_{\nu}$/d.o.f.&  Disk Flux  \\
&& \textbf{or} (C-stat)&($10^{-14}$ erg cm$^{-2}$ s $^{-1})$&(keV)&($10^{-4}$)& \textbf{or} (C-stat)&($10^{-14}$ erg cm$^{-2}$ s $^{-1})$\\ \hline  \hline
 322 (2000-03-19) &  1.9($\pm 0.5)$ &(220.21/38)&4.3 ($^{+1.7}_{-1.2})$&1.20 ($^{+0.90}_{-0.42})$ & 10.6 ($^{+21.8}_{-10.6})$& (30.64/38)& 4.2 ($^{+2.0}_{-1.3})$ \\ 
321 (2000-06-12) &1.6  ($\pm 0.2)$ & 0.64/7 & 5.3 ($\pm 0.9)$ &0.98  ($^{+0.38}_{-0.24})$  &20.9  ($^{+35.9}_{-14.0})$ & 1.32/7 & 3.7 ($^{+1.0}_{-0.8})$\\
8095 (2008-02-23) & 1.4 ($\pm 0.6$) &(25.54/22)& 3.8 ($^{+2.6}_{-1.6})$ &1.2  ($^{+1.9}_{-0.5})$ & $\leq$ 10.6$^{**}$ &(24.54/22)& 3.0 ($^{+2.4}_{-1.3})$ \\
 11274 (2010-02-27) &1.5 ($\pm 0.2)$  &0.82/7&  4.4($\pm 0.9$) &1.52 ($^{+0.63}_{-0.39})$& 3.7 ($^{+0.6}_{-0.3})$& 1.69/7& 3.8 ($\pm$ 0.9) \\ 
 %2010-11-20 &---\footnote{fit with fixed \textbf{power law} of 1.6} &---&  N/A&  2.61 ($^{+17.02}_{-12.99})$ $\times$ $10^{-15}$  \\ 
12889 (2011-02-14) &1.5 ($\pm 0.1$) & 1.42/25 &3.63$\pm 0.4$ )&1.28 ($^{+0.22}_{-0.18})$ & 5.9 ($^{+4.0}_{-2.5})$&1.48/25& 3.0 ($\pm$ 0.4) \\ 
12888 (2011-02-21) & 1.5 ($\pm 0.1$)  & 1.20/33 & 4.0 ($\pm 0.4)$&1.39 ($^{+0.25}_{-0.19})$& 4.7 ($^{+3.3}_{-2.0})$&1.82/33& 3.3 ($\pm$ 0.4) \\ 
 16260 (2014-08-04)   & 1.5 ($\pm 0.4$ ) & (41.84/48) & 2.5($^{+0.9}_{-0.69})$&1.4 ($^{+1.2}_{-0.5})$&2.7 ($^{+7.9}_{-2.7})$&(43.73/48)& 2.2 ($^{+1.0}_{-0.6})$\\ 
 16261 (2015-02-24) &  1.5 ($\pm 0.5$)& (43.75/58)  & 4.1($^{+1.2}_{-0.9})$& 1.46($^{+0.88}_{-0.42})$&$\leq$ 13.8$^{**}$ &(44.97/58)& 3.6 ($^{+1.2}_{-0.9})$\\ 
 16262 (2016-04-30)  & 1.5 ($\pm 0.4)$ & (48.88/50) & 3.5($_{-0.9}^{+1.2}$)&1.5 ($^{+1.2}_{-0.5})$&$\leq$ 15.5$^{**}$&(48.94/50)& 3.0 ($^{+1.3}_{-0.9})$\\ 
\hline
\end{tabular}
\end{table*}
\begin{table*}
\centering
\caption{F-test probability values for single component versus two component models for GC ULXs in NGC~4472 with over 100 source counts. We compare statistics between \texttt{tbabs*(diskbb+pegpwrlw)} and \texttt{tbabs*diskbb} only in columns titled ``Disk", and \texttt{tbabs*(diskbb+pegpwrlw)}  with \texttt{tbabs*pegpwrlw} only in columns titled ``PL". Blank table entries are where the source had fewer than 100 counts in a given observation.  }
\label{4472_ftest}
\begin{tabular}{l|ll|ll|ll|ll|ll}
\hline
& GCU1 &&GCU2&&GCU3&&GCU4 \\
\hline
ObsID & Disk  & PL & Disk  &  PL  & Disk  & PL & Disk  & PL   \\\hline
321   & 1      & 0.31   & 1& 0.06   & 0.01 & 0.99    & 0.15 & 0.99          \\  
11274 & 0.07       & 0.06     &&     &&        &&  \\
12889 & 0.18      & 5.65e-8    &0.10 & 0.17    & 2.31e-3 & 4.07e-3  & 0.06 & 0.01                        \\ 
12888 & 0.14    & 1.37e-7    &0.36 & 5.83e-4   &0.10& 6.01e-6& 1.35e-3 &1           \\ \hline 
\end{tabular}
\end{table*}

\subsubsection{NGC 4649 GC ULXs}

\begin{figure}

\includegraphics[scale=0.45]{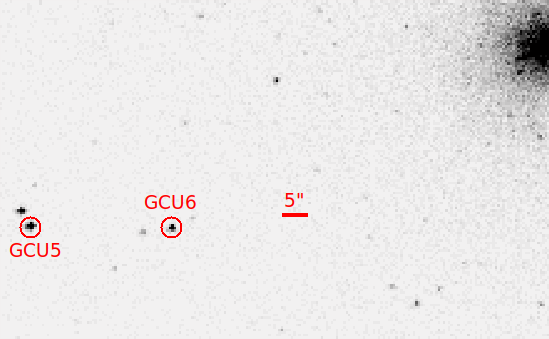}

\caption{X-ray image of NGC 4649 (ObsID 12976, filtered to 0.5-8.0~keV) with regions for GCU5 \& GCU6 overlaid. } 
\label{Fig:img4649}
\end{figure}

NGC 4649 hosts two GC ULXs, CXOUJ1243469+113234 (hereafter $\RS$) \citep{2012ApJ...760..135R} and CXOU1243445+113150 (hereafter $\CC$)  (see Figure \ref{Fig:img4649} for a \textit{Chandra} image of the source locations.).
\begin{table}
\centering
\caption{Chandra Observations of NGC 4649, with raw source counts (0.5-8.0 keV)  for $\RS$ and $\CC$.}
\label{obs4649}
\begin{tabular}{lclclclclc}
\hline
\hline
ObsID & Date       & ObsLen  & GCU5 & GCU6\\ 
&& (ks) & Cts & Cts \\ \hline
785   & 2000-04-20 & 38.11    &65&  135 \\ 
8182  & 2007-01-30 & 52.37     &468& 196 \\ 
8507  & 2007-02-01 & 17.52     &111& 45 \\ 
12976 & 2011-02-24 & 101.04     &773& 265\\ 
12975 & 2011-08-08 & 84.93      &505& 260\\ 
14328 & 2011-08-12 & 13.97      &84& 39\\ \hline
\end{tabular}
\end{table}
\subsubsubsection{CXOUJ1243469+113234 (GCU5)}

As noted in \citet{2012ApJ...760..135R}, $\RS$ shows better fit statistics for the single power law component. ObsID 12975 is an exception here; the single disk component had much better fit statistics than the single power law. However, \citet{2012ApJ...760..135R} found that this observation had a second intrinsic absorbing column that was significant. We do not account for such a component here, but it could explain why in this particular instance, a disk was a better fit than a single power law, when a power law has typically a better fit than a disk in previous observations. The F-test values for comparison between a single component model and a two component model are listed in Table \ref{4649fits} and the best fit values for either single component models are listed in Table \ref{4649fits_tbabs}.

%\begin{table}
%\centering
%\caption{F-test probability values for single component versus two component models for observations of %$\RS$  with over 100 source counts.  We compare statistics between \texttt{tbabs*(diskbb+pegpwrlw)} and %\texttt{tbabs*diskbb} only in columns titled "Disk", and \texttt{tbabs*(diskbb+pegpwrlw)}  with %\texttt{tbabs*pegpwrlw} only in columns titled "PL".}
%\label{4649_ftest}
%\begin{tabular}{lllll}
%\hline
%\hline
%ObsID & Disk  &  PL  \\\hline
%8182  & 0.09   & 0.31                               \\
%12976 & 2.72 $\times 10^{-3}$          & 0.19                   \\
%12975 & 0.99  & 6.41 $\times 10^{-3}$    \\ \hline
%\end{tabular}
%\end{table}

\subsubsubsection{CXOU1243445+113150 (GCU6)}
$\CC$  is ambiguous as to whether a single disk model or a single power law model is a better fit. See Table \ref{4649fits} for a comparison of the single disk versus power law model.The F-test values  for $\CC$ (Table \ref{4649_ftest}) indicate that the two component model is not a better fit to the data. We find that this source has a significant (non-zero) second absorbing column for ObsIDs 8182 and 12975. When comparing these observations to a disk model with a best fit second absorbing column that is consistent with zero, we find that it is  statistically ambiguous as to which model is a better fit. We present these fits in Table \ref{4649fits_tbabs}. 

\begin{table*}
\caption{\textit{Chandra} Fit Parameters and Fluxes (0.5-8 keV) for spectral best fit single-component models, \texttt{tbabs*diskbb} and \texttt{tbabs*pegpwrlw}  of of GC ULXs in NGC 4649. (Hydrogen column density ($N_H$) frozen to 2.0 $\times 10^{20}$  cm$^{-2}$). Fit parameters marked with $^{**}$ either encountered an \textsc{xspec} error when computing a lower bound, or had a lower bound consistent with zero, and are presented as an upper limit. Lower count observations fit with C-stat have their statistics presented in parentheses. All fluxes shown are unabsorbed. }
\label{4649fits}
\begin{tabular}{|l|clc|lclc|}
\hline
\hline
\toprule &&& $\RS$ \\ \hline
  &       \multicolumn{3}{c}{\textsc{tbabs*pegpwrlw}} & \multicolumn{3}{|c}{\textsc{tbabs*diskbb}} \\ \hline
\midrule
 ObsID/Date  & $\Gamma$ &   $\chi^2_{\nu}$/d.o.f.& PL Flux  & $T_{in}$ &  Disk Norm &  $\chi^2_{\nu}$/d.o.f. &  Disk Flux  \\
&& \textbf{or} (C-stat)&($10^{-14}$ erg cm$^{-2}$ s $^{-1})$&(keV)&($10^{-4}$)& \textbf{or} (C-stat)&($10^{-14}$ erg cm$^{-2}$ s $^{-1})$\\ \hline  \hline

 785 (2000-04-20) &1.6 ($\pm$0.3)&(61.54/59)&1.3  ($^{+0.5}_{-0.4})$ &  1.08 ($^{+0.60}_{-0.30})$ & $\leq$ 12.5$^{**}$&  (65.99/59)& 1.1 ($^{+0.5}_{-0.3})$ \\ 
8182 (2007-01-30)  &1.2 ($\pm$ 0.1)&1.47/21&9.3 ($\pm$ 1.1)& 2.02($^{+0.50}_{-0.36})$ &2.9($^{2.5}_{-1.5})$& 1.68/21 & 8.6 ($\pm 1.2)$   \\
%2008-02-23 & ---&---&N/A& $\leq$1.61 $\times 10^{-14}$  \\
 8507 (2007-02-01)  &1.5 ($\pm$ 0.3)&2.52/4&5.3 ($^{+1.3}_{-1.2})$&1.27 ($^{+0.45}_{-0.31})$   &9.2($^{+14.4}_{-5.9})$&2.00/4&  4.6($\pm 1.1)$ \\ 
 %2010-11-20 &---\footnote{fit with fixed \textbf{power law} of 1.6} &---&  N/A&  2.61 ($^{+17.02}_{-12.99})$ $\times$ $10^{-15}$  \\
 12976 (2011-02-24) &1.6 ($\pm$ 0.1)&1.50/36&6.2 ($\pm$ 0.6)& 1.09 ($\pm 0.13)$&19.4($^{+9.9}_{-6.6})$&1.93/36 &5.3 ($\pm 0.5$ ) \\ 
 12975 (2011-08-08) &1.1 ($\pm$ 0.1)&1.20/23&7.0 ($\pm$ 0.7)& 2.18 ($^{+0.54}_{-0.37})$  & 1.6 ($^{+1.3}_{-0.8})$ &0.74/23 & 6.2($\pm 0.8)$\\ 

 %2014-08-04   &---\footnote{fit with fixed \textbf{power law} of 1.6}   &--- & N/A & 1.84($^{+13.20}_{-9.56})$$\times$$10^{-15}$\\ 
14328 (2011-08-12) &1.2 ($\pm$ 0.3)&(48.53/77)&7.02($^{+2.05}_{-1.60})$& 2.1  ($^{+2.1}_{-0.7})$&$\leq$ 7.20$^{**}$& (49.51/76)  & 6.4($^{+2.3}_{-1.7})$\\

\hline
\toprule 
 &&& $\CC$\\ \hline
\midrule
 ObsID/Date  & $\Gamma$ &   $\chi^2_{\nu}$/d.o.f.  & PL Flux  & $T_{in}$ &  Disk Norm &  $\chi^2_{\nu}$/d.o.f.  &  Disk Flux  \\
&& \textbf{or} (C-stat)&($10^{-14}$ erg cm$^{-2}$ s $^{-1})$&(keV)&($10^{-4}$)& \textbf{or} (C-stat)&($10^{-14}$ erg cm$^{-2}$ s $^{-1})$\\ \hline  \hline
 785 (2000-04-20) & 1.3 ($\pm$0.2)&0.95/5&3.2 ($^{+0.9}_{-0.8})$ & 1.32 ($^{+0.94}_{-0.43})$ &$\leq$ 10.3$^{**}$&1.71/5 &2.3 ($^{+1.1}_{-0.7})$ \\ 
8182 (2007-01-30)  &1.3 ($\pm$ 0.2)&1.41/8&3.7 ($\pm$ 0.6)& 1.28($^{+0.34}_{-0.25})$ &5.5 ($^{6.2}_{-3.1})$& 0.95/8 & 2.9 ($\pm 0.6)$  \\
%2008-02-23 & ---&---&N/A& $\leq$1.61 $\times 10^{-14}$  \\
8507 (2007-02-01) &1.5 ($\pm$ 0.4)&(35.81/43)&2.4($^{+1.1}_{-0.7})$ &1.27 ($^{+0.92}_{-0.41})$   &$\leq$ 13.6$^{**}$ &(43.23/44)&  2.0 ($^{+0.9}_{-0.6})$   \\ 
 %2010-11-20 &---\footnote{fit with fixed \textbf{power law} of 1.6} &---&  N/A&  2.61 ($^{+17.02}_{-12.99})$ $\times$ $10^{-15}$  \\ 

 12976 (2011-02-24)& 1.6 ($\pm$ 0.2)&1.34/12&2.2 ($\pm$ 0.3)&1.12 ($^{+0.31}_{-0.23})$  & 5.7 ($^{+7.2}_{-3.3})$& 2.04/12 & 5.3($\pm 0.3)$\\ 
  12975 (2011-08-08) &1.1 ($\pm$ 0.2)&1.96/12&3.6 ($\pm$0.5)&2.15 ($^{+0.90}_{-0.51})$&0.9 ($^{+1.2}_{-0.6})$& 1.67/12 &3.2($\pm 0.6$) \\ 
 %2014-08-04   &---\footnote{fit with fixed \textbf{power law} of 1.6}   &--- & N/A & 1.84($^{+13.20}_{-9.56})$$\times$$10^{-15}$\\ 
14328 (2011-08-12)  &2.0 ($\pm$ 0.5)&(27.63/37)&2.1 ($\pm$ 0.2)& 0.65  ($^{+0.32}_{-0.15})$& $\leq$ 64.5$^{**}$& (31.03/37)  & 1.6 ($^{+0.6}_{-0.4})$\\ 

\hline
\end{tabular}
\end{table*}
\begin{table}
\centering
\caption{F-test probability values for single component versus two component models for  GC ULXs in NGC~4649 with over 100 source counts.  We compare statistics between \texttt{tbabs*(diskbb+pegpwrlw)} and \texttt{tbabs*diskbb} only in columns titled "Disk", and \texttt{tbabs*(diskbb+pegpwrlw)}  with \texttt{tbabs*pegpwrlw} only in columns titled "PL".}
\label{4649_ftest}
\begin{tabular}{lllll}
\hline
& GCU5&&GCU6& \\
\hline
ObsID & Disk  &  PL & Disk &  PL  \\\hline
785 &-&- & 0.33  &  0.80 \\
8182  & 0.09   & 0.31    & 0.65&        0.20                             \\
12976 & 2.72 $\times 10^{-3}$          & 0.19     &0.12     &     1                \\
12975 & 0.99  & 6.41 $\times 10^{-3}$   & 1 &      0.47  \\ \hline
\end{tabular}
\end{table}

\begin{table*}
\caption{\textit{Chandra} Fit Parameters and Fluxes (0.5-8.0~keV) for spectral  best fit values  of GCU6 in NGC 4649 for \texttt{tbabs*tbabs*pegpwrlw} where $N_H$ was not consistent with zero, and $\chi^2$/d.o.f. for \texttt{tbabs*tbabs*diskbb},where the best fit $N_H$ was consistent with zero. (Hydrogen column density ($N_H$) frozen to 2.2 $\times 10^{20}$  cm$^{-2}$. All fluxes shown are unabsorbed. }
\label{4649fits_tbabs}
\begin{tabular}{|l|cclc|lclc|}
\hline
\hline
\toprule 
  \hline
\midrule
 ObsID/Date &$N_H$ & $\Gamma$ &  PL  $\chi^2_{\nu}$/d.o.f.  & PL Flux &  Disk $\chi^2_{\nu}$/d.o.f.  \\
 &($ 10^{20}$  cm$^{-2}$)&&&($10^{-14}$ erg cm$^{-2}$ s $^{-1}$)&&\\
\hline
\hline
8182 (2007-01-30) &0.17 $^{+0.17}_{-0.13}$ &1.8 ($\pm$ 0.4)&0.84/7&3.7 ($\pm$ 0.6)&  1.1/7 \\
%2008-02-23 & ---&---&N/A& $\leq$1.61 $\times 10^{-14}$  \\

  12975 (2011-08-08) &0.25$^{+0.25} _{-0.23}$&1.5 ($\pm$ 0.4)&1.85/11&3.7 ($\pm$0.5)& 1.82/11  \\ 
 %2014-08-04   &---\footnote{fit with fixed power law of 1.6}   &--- & N/A & 1.84($^{+13.20}_{-9.56})$$\times$$10^{-15}$\\ 

\hline
\end{tabular}
\end{table*}
\subsubsection{NGC~1399 GC ULXs}

\begin{figure}
\includegraphics[scale=0.5]{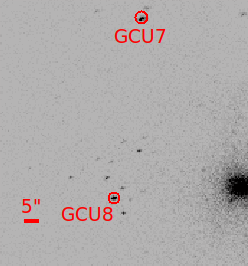}

\caption{X-ray image of NGC 1399 (ObsID 319, filtered to 0.5-8.0 keV) with regions for GCU7 \& GCU8 overlaid. } 
\label{fig:img1399}
\end{figure}

NGC 1399 hosts three GC ULXs. One, CXOKMZJ033831.7$-$353058 \citep{2010ApJ...721..323S} has faded beyond detection by 2005, and we do not study it here. CXOJ0338318-352604, (hereafter $\irwin$)  has been previously studied by \citet{irwin2010}, while the third, CXOU0338326-35270567 (hereafter $\BC$) is bright but has not been previously extensively studied in X-ray. We present new analysis on both old and new data (Table \ref{obs1399}). see Figure \ref{fig:img1399} for a \textit{Chandra} image of the source locations and any nearby sources. 

%The cluster associated with $\irwin$ is z-magnitude 20.7, and g-z of 1.98 \citep{Paolillo11}.  The cluster hosting $\BC$ is z-magnitude 19.9 with g-z of 2.24 \citep{Paolillo11}.

\begin{table}
\centering
\caption{Chandra Observations of NGC 1399, with raw source counts (0.5-8.0 keV) for $\irwin$   and $\BC$.}
\label{obs1399}
\begin{tabular}{lclclclclc}
\hline
\hline
ObsID & Date       & ObsLen  & GCU7  & GCU8\\ 
&& (ks) &Cts & Cts\\ \hline
320   & 1999-10-18 & 3.38       &12& 23  \\ 
319   & 2000-01-18 & 56.04    &448&  629 \\ 
239   & 2000-01-19 & 3.60     &14& 22  \\ 
240   & 2000-06-16 & 43.53    &12& 44   \\ 
2389  & 2001-05-08 & 14.67    &13& 10  \\ 
4172  & 2003-05-26 & 44.50    &114&294   \\ 
9530  & 2008-06-08 & 59.35   &248& 366    \\ 
14527 & 2013-07-01 & 27.79   &136& 230   \\ 
16639 & 2014-10-12 & 29.67   &171& 162   \\ 
14529 & 2015-11-06 & 31.62   &132& 226   \\ \hline
\end{tabular}
\end{table}

\subsubsubsection{CXOJ0338318-352604 (GCU7)}
$\irwin$ shows better fit statistics for the single component disk model (Table \ref{1399fits}). The F-test values indicate that the two component model is not necessary, except perhaps for ObsID 319 (Table \ref{1399_ftest}). However, the two component model has a best fit power law index of 5.7. Fitting this source with a second absorption component plus power law gives an absorption that is consistent with zero in almost all cases, and for all fits of \texttt{tbabs*tbabs*pegpwrlw}, the  power law index is between 3-4, which implies that the spectrum is very soft and unlikely to be fit by a power law model of any sort.

\citet{clausen12} model this system as the tidal disruption of a horizontal giant branch star by an intermediate mass black hole, while \citet{2011MNRAS.410L..32M} model this as a R Corona Borealis star illuminated by a bright X-ray source. Neither scenario is ruled out by the observed shallow decline of the $L_X$. Other binary system models might also be consistent with these data.

%\textbf{\citet{clausen12} model this system as the tidal disruption of a horizontal giant branch star by an intermediate mass black hole. The shallow decline of the $L_X$ in the model is not inconsistent with the current data set, however, the R Corona Borealis scenario proposed by \citet{2011MNRAS.410L..32M} is 

%\begin{table}
%\centering
%\caption{F-test probability values for single component versus two component models for observations of $\irwin$  with over 100 source counts.  We compare statistics between \texttt{tbabs*(diskbb+pegpwrlw)} and \texttt{tbabs*diskbb} only in columns titled "Disk", and \texttt{tbabs*(diskbb+pegpwrlw)}  with \texttt{tbabs*pegpwrlw} only in columns titled "PL".}
%\label{1399_ftest}
%\begin{tabular}{lllll}
%\hline
%\hline
%ObsID & Disk  & PL \\\hline
%319   & 0.08   & 7.63 $\times 10^{-4}$                             \\
%4172  & 0.22     & 0.46                        \\
%9530  & 0.56   & 0.04                        \\
%14527 & 0.54   & 0.11                           \\
%16639 & 0.11      & 0.03                       \\
%14529 & 0.54  & 0.37     \\ \hline
%\end{tabular}
%\end{table}

\subsubsubsection{CXOU0338326-35270567 (GCU8)}
We find that there is no evidence for a second absorbing column of any sort for this source.The F-test values for $\BC$ rule out a two component model (Table \ref{1399_ftest}), and the statistics indicate that a single power law component is the best fit model for this source. (Table \ref{1399fits}). Note that the best fit values of the disk norms for $\BC$ are upper limits, even in the higher count observations binned by 20 and fit with $\chi^2$ statistics, most likely because the single disk fit for this source was not a good fit to the point where \textsc{xspec} had difficulty fitting the normalisations to these spectra.  

GCU8 was marginally detected in ObsID 2389. We used the WebPIMMS tool to estimate an upper limit on the unabsorbed flux using a count rate of 6.8$\times 10^{-4}$ ct/s, and a fixed powerlaw index of 1.7, with the $N_H$ fixed to frozen to 1.34 $\times 10^{20}$  cm$^{-2}$ .

\begin{table*}
\begin{minipage}{\textwidth}
\caption{\textit{Chandra} Fit Parameters and Fluxes (0.5-8 keV) for spectral best fit single-component models, \texttt{tbabs*diskbb} and \texttt{tbabs*pegpwrlw}  of GC ULXs in NGC 1399. Hydrogen column density ($N_H$) frozen to 1.34 $\times 10^{20}$  cm$^{-2}$). Fit parameters marked with $^{**}$ either encountered an \textsc{xspec} error when computing a lower bound, or had a lower bound consistent with zero, and are presented as an upper limit. Lower count observations fit with C-stat have their statistics presented in parentheses. All fluxes shown are unabsorbed.}
\label{1399fits}
\begin{tabular}{|l|clc|lclc|}
\hline
\hline
\toprule &&& $\irwin$ \\ \hline
  &       \multicolumn{3}{c}{\textsc{tbabs*pegpwrlw}} & \multicolumn{3}{|c}{\textsc{tbabs*diskbb}} \\ \hline
\midrule
 ObsID/Date  & $\Gamma$ &   $\chi^2_{\nu}$/d.o.f. & PL Flux & $T_{in}$ &  Disk Norm &  $\chi^2_{\nu}$/d.o.f.  &  Disk Flux \\
&& \textbf{or} (C-stat)&($10^{-14}$ erg cm$^{-2}$ s $^{-1})$&(keV)&($10^{-2}$)& \textbf{or} (C-stat)&($10^{-14}$ erg cm$^{-2}$ s $^{-1})$\\ \hline  \hline

320 (1999-10-18) &3.1 ($^{+1.0}_{-0.9})$&(10.91/12)&2.1 ($^{+1.5}_{-1.0})$&0.28 ($^{+0.19}_{-0.09})$ & $\leq$ 46.8$^{**}$&  (8.50/12)&4.8 ($^{+4.0}_{-3.2})$ \\
319 (2000-01-18) &2.4 ($\pm$ 0.1)&8.32/24&2.9 ($\pm$ 0.3)& 0.39  ($\pm 0.04)$ &7.3 ($^{+3.5}_{-2.4})$ &1.70/24&  2.5($^{+0.2}_{-0.3})$  \\
 239 (2000-01-19) &3.1 ($^{+1.1}_{-0.9})$ &(14.53/13)&2.5 ($^{+1.7}_{-1.1})$ &  0.27 ($^{+0.18}_{-0.10})$ & $\leq$ 288.4$^{**}$& (12.36/13) &2.3 ($^{+1.3}_{-1.0})$ \\ 
240 (2000-06-16) &2.7 ($^{+0.8}_{-0.7})$ &(24.99/17)& 1.6($^{+0.9}_{-0.7})$& 0.52  ($^{+0.29}_{-0.18})$ &1.3 ($^{+1.6}_{-0.7})$& (22.28/17) & 1.5($^{+0.8}_{-6.2})$ \\
 
 %2010-11-20 &---\footnote{fit with fixed \textbf{power law} of 1.6} &---&  N/A&  2.61 ($^{+17.02}_{-12.99})$ $\times$ $10^{-15}$  \\ 
2389 (2001-05-08) &3.8 ($^{+1.1}_{-1.0})$ &(10.30/12)&4.1  ($^{+2.4}_{-1.8})$  & 0.21 ($^{+0.07}_{-0.11})$  & $\leq$ 1272$^{**}$& (8.94/12)& 4.3($^{+2.5}_{-1.8})$  \\ 
4172 (2003-05-26) &2.9 ($\pm 0.3$)&0.76/5& 2.1 ($\pm$ 0.3)& 0.36 ($^{+0.08}_{-0.06}$) & 7.7($^{+11.9}_{-4.7})$ & 0.96/5 & 1.7($\pm 0.3)$\\ 
 %2014-08-04   &---\footnote{fit with fixed \textbf{power law} of 1.6}   &--- & N/A & 1.84($^{+13.20}_{-9.56})$$\times$$10^{-15}$\\ 
9530 (2008-06-08) &2.6 ($\pm$ 0.2)&2.64/12&2.4 ($\pm$ 0.3)& 0.34  ($\pm 0.04$) & 11.9($^{+8.2}_{-5.0})$& 1.25/12  & 2.0($\pm 0.2$)\\ 
14527 (2013-07-01) &2.7 ($\pm$ 0.3)&3.71/6&3.2 ($\pm 0.5$)& 0.39  ($^{+0.08}_{-0.06}$) & 7.9($^{+9.2}_{-4.4})$&   0.98/6 &2.7($\pm 0.4$)\\ 
16639 (2014-10-12) &2.7 ($\pm 0.2$)&1.99/7&4.0 ($\pm$ 0.6)& 0.43  ($\pm 0.07$) &6.6($^{+7.1}_{-3.3})$&  0.94/7 &3.4($\pm 0.5$)\\ 
14529 (2015-11-06)  &3.0 ($\pm 0.3$)&3.23/5& 3.2 ($\pm 0.5$)& 0.35  ($\pm 0.07$) &14.2($^{+21.3}_{-8.5})$&   1.65/5 &2.7($\pm 0.4$)\\

\hline

\toprule
  &      && $\BC$\\ \hline
\midrule
 ObsID/Date  & $\Gamma$ &   $\chi^2_{\nu}$/d.o.f.  & PL Flux  & $T_{in}$ &  Disk Norm &  $\chi^2_{\nu}$/d.o.f.  &  Disk Flux \\
&& \textbf{or} (C-stat)&($10^{-14}$ erg cm$^{-2}$ s $^{-1})$&(keV)&($10^{-4}$)& \textbf{or} (C-stat)&($10^{-14}$ erg cm$^{-2}$ s $^{-1})$\\ \hline  \hline
320 (1999-10-18) &1.3 ($\pm 0.7$)&(18.87/20)&8.8 ($^{+1.7}_{-3.9})$&$\leq$ 1.8 & $\leq$ 4.09$^{**}$&  (19.54/18)&7.8 ($^{+9.2}_{-3.9})$ \\
319 (2000-01-18) & 1.4  ($\pm 0.1)$ &1.37/31&  8.3 ($\pm 0.9)$ &1.27($^{+0.21}_{-0.17})$     &13.5 ($^{+5.2}_{-8.2})$  &1.84/31& 6.7 ($\pm$ 0.9) \\
240 (2000-06-16) &1.5 ($\pm 0.5)$ &(51.46/48)&  5.9 ($\pm$ 1.7)& 1.8 ($^{+1.3}_{-0.5})$ &$\leq$ 3.14$^{**}$ & (53.01/48) & 5.5($^{+1.7}_{-1.3})$ \\
 2389 (2001-05-08) & -& -& -& -&- &- & $\leq$  0.5 \footnote{Upper limit calculated using \url{http://cxc.harvard.edu/toolkit/pimms.jsp}} \\
 %2010-11-20 &---\footnote{fit with fixed \textbf{power law} of 1.6} &---&  N/A&  2.61 ($^{+17.02}_{-12.99})$ $\times$ $10^{-15}$  \\ 

 %2010-11-20 &---\footnote{fit with fixed \textbf{power law} of 1.6} &---&  N/A&  2.61 ($^{+17.02}_{-12.99})$ $\times$ $10^{-15}$  \\ 
4172 (2003-05-26) & 1.1 ($\pm 0.2$)  & 1.24/12& 9.5($\pm 1.4)$&2.6 ($^{+2.0}_{-0.8})$ & $\leq$ 3.78$^{**}$ &1.74/12& 8.7 ($^{+1.8}_{-1.5})$ \\ 

 %2014-08-04   &---\footnote{fit with fixed \textbf{power law} of 1.6}   &--- & N/A & 1.84($^{+13.20}_{-9.56})$$\times$$10^{-15}$\\ 
9530 (2008-06-08) & 1.5  ($\pm 0.2$) & 0.66/16 & 5.6($\pm 0.7$&1.28 ($\pm$ 0.30)&8.7 ($^{+8.3}_{-4.4})$ &1.25/16&4.5 ($\pm$ 0.8)\\ 
14527 (2013-07-01) & 1.3  ($\pm 0.2$) &  0.42/9 &9.7($\pm 1.7$)&1.36($^{+0.47}_{-0.29})$ &11.0 ($^{+14.1}_{-6.9})$ &0.28/9&7.2 ($^{+1.6}_{-1.3})$\\ 
16639 (2014-10-12) & 1.1 ($\pm 0.3$) &  0.78/6 & 8.0$^{+1.9}_{-1.7})$&2.0($^{+2.9}_{-0.7})$&$\leq$ 9.09$^{**}$&0.96/6& 6.3 ($^{+2.7}_{-1.7})$\\ 
14529 (2015-11-06)  & 1.2 ($\pm 0.2$) &   0.52/9 & 10.2($^{+1.9}_{-1.7})$ & 2.0($^{+1.2}_{-0.5})$& $\leq$ 9.46$^{**}$ &0.67/9& 8.4 ($^{+2.2}_{-1.8})$\\

\hline
\end{tabular}
\end{minipage}
\end{table*}

\begin{table}
\centering
\caption{F-test probability values for single component versus two component models for GC ULXs in NGC~1399 with over 100 source counts.  We compare statistics between \texttt{tbabs*(diskbb+pegpwrlw)} and \texttt{tbabs*diskbb} only in columns titled ``Disk", and \texttt{tbabs*(diskbb+pegpwrlw)}  with \texttt{tbabs*pegpwrlw} only in columns titled ``PL".}
\label{1399_ftest}
\begin{tabular}{l|c|c|c|c}
\hline
& GCU7 && GCU8 &\\
\hline
ObsID & Disk  & PL & Disk & PL \\\hline
319   & 0.08   & 7.63 $\times 10^{-4}$   & 6.14 $\times 10^{-3}$ &  0.44                             \\
4172  & 0.22     & 0.46  & 0.19&         1                          \\
9530  & 0.56   & 0.04    &  0.01&       1                       \\
14527 & 0.54   & 0.11     & 0.51&      0.12                       \\
16639 & 0.11      & 0.03   & 0.65&          1                       \\
14529 & 0.54  & 0.37     &  0.42&  0.99 \\ \hline
\end{tabular}
\end{table}

\subsection{Long and Short-term X-ray Variability}
\label{appendix-var}
	RZ2109 has been shown to vary both long and short term (see \citealt[][for example]{2007Natur.445..183M, 2010ApJ...721..323S, 2018arXiv180601848D}), which is leading evidence for its compact object being a black hole accretor. Other GC ULXs show some significant long-term variability, and quantifying the variability of these sources on either long or short timescales can shed light on the nature of the objects that make up these ULX systems. 
    
\subsubsection{Long-Term Variability}  
\begin{figure}
\includegraphics[scale=0.4]{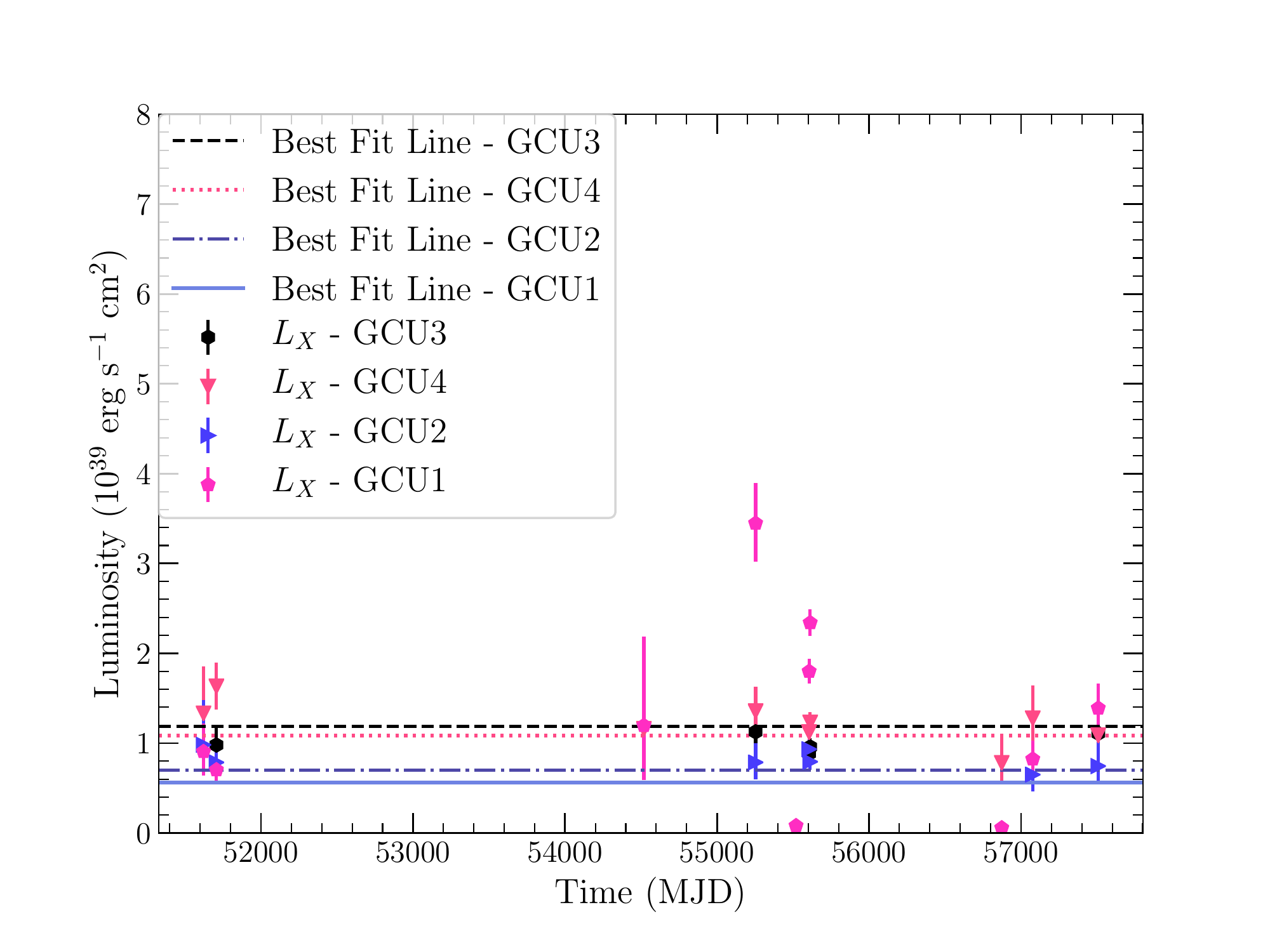}
\includegraphics[scale=0.4]{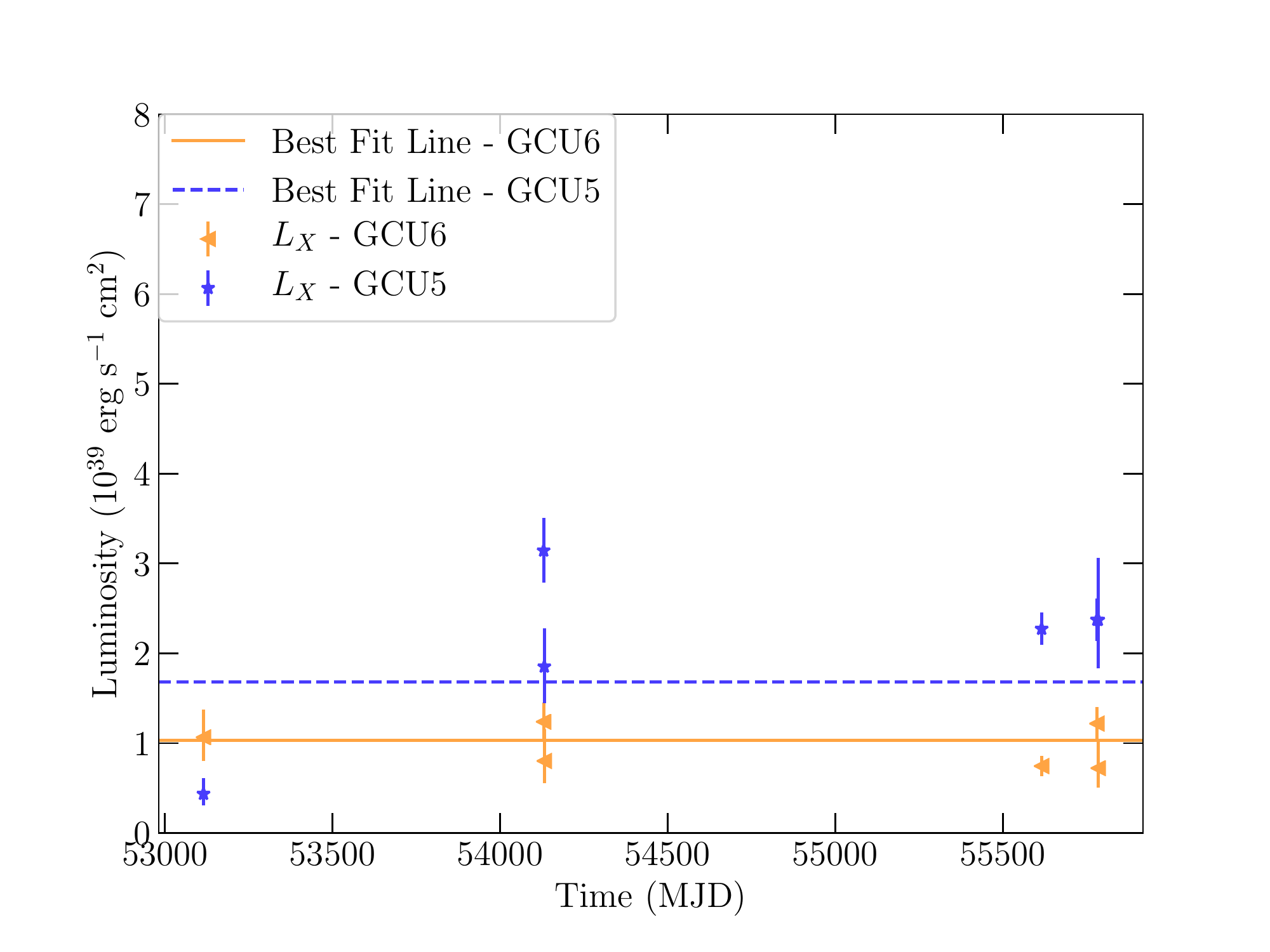}
\includegraphics[scale=0.4]{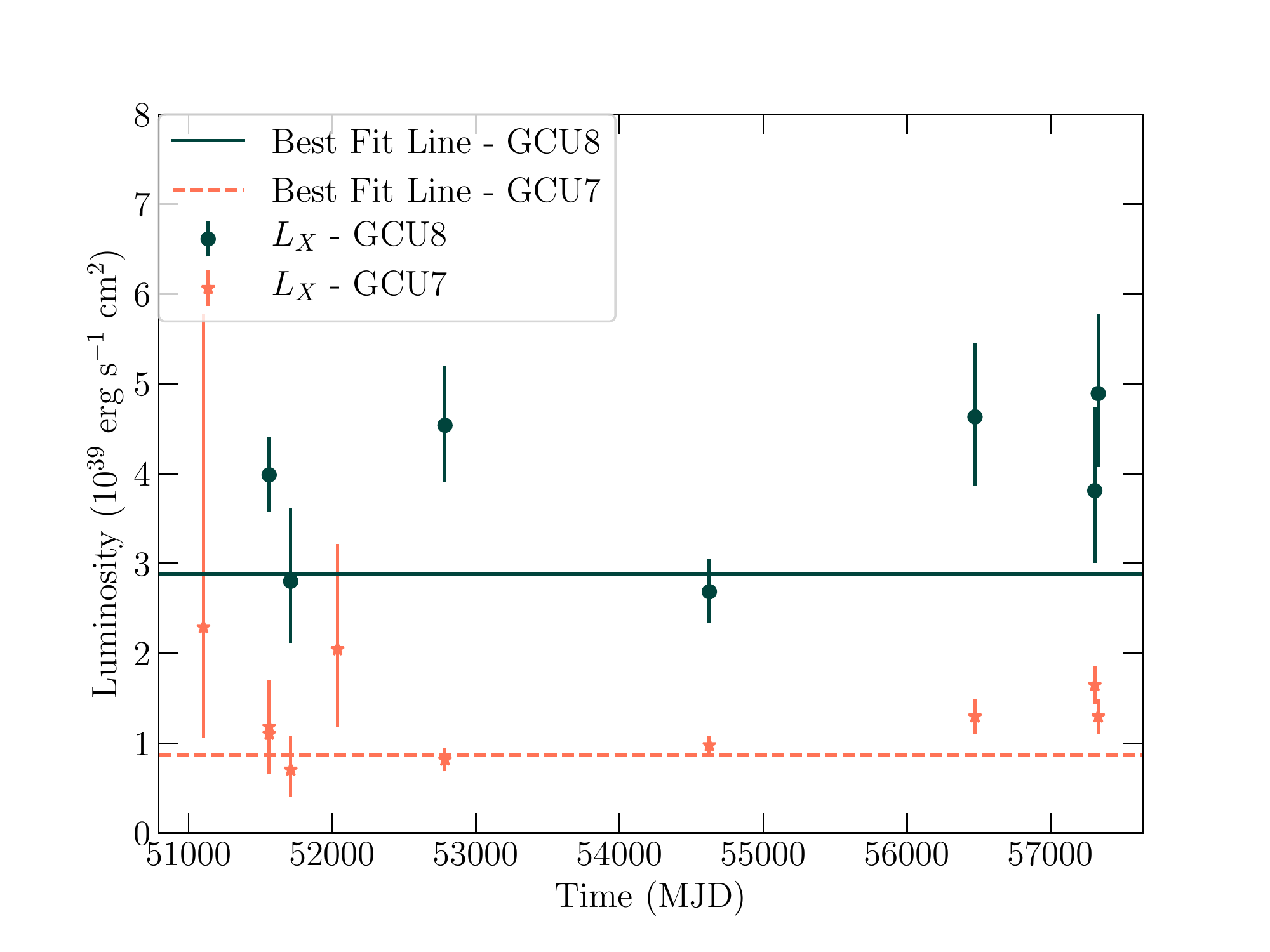}

\caption{Upper: $L_X$ vs. time for NGC 4472 GC ULXs with mean luminosity (data from Tables \ref{4472fits}). Middle: $L_X$ vs. time for NGC 4649 GC ULXs with mean luminosity (data from Tables \ref{4649fits}).  Lower:  $L_X$ vs. time for NGC 1399 GC ULXs with mean luminosity (data from Tables \ref{1399fits} ).}  
\label{fig:all_ltvar}
\end{figure}

   We found the average luminosity comparing the data to a range of 5000 luminosities drawn from the lowest $L_X$ to the highest for the source and computing the reduced $\chi^2$ in each case. We plot the luminosity with the $\chi^2$ closest to 1.0 as the mean luminosity.  This is plotted for all the sources in Figure \ref{fig:all_ltvar}.

To quantify the variability of these sources, we  do a $\chi^2$ minimisation fit of the data using \textsc{scipy}\footnote{https://docs.scipy.org/doc/scipy/reference/generated/scipy.stats.chisquare.html} with a model of constant luminosity, using the best fit luminosity from above as the mean value.  We find that GCU2, GCU3, GCU4 and GCU6 have $\chi^2$ values less than 1 (e.g., no evidence for long-term variability), while GCU1, GCU5, GCU6 and GCU7 have very large $\chi^2$ values, indicating that they are much more variable. These values are presented in Table \ref{chi2}. 
\begin{table*}
\centering
\caption{$\chi^2$ values comparing a model of constant luminosity to the luminosity of the GC ULXs in this sample over time.  }
\label{chi2}
\begin{tabular}{|l|l|l|l|l|l|l|l|l|}
\hline
Source   & GCU1  & GCU2 & GCU3 & GCU4 & GCU5 & GCU6 & GCU7 & GCU8 \\ \hline
$\chi^2$ & 26.33 & 0.22 & 0.17 & 0.55 & 3.00 & 0.30 & 5.22 & 4.12 \\ \hline
\end{tabular}
\end{table*}

%\begin{figure}

%\includegraphics[scale=0.4]{variability_GCU3.pdf}

%\includegraphics[scale=0.4]{variability_GCU4.pdf}

%\caption{Upper: $L_X$ vs. time for $\AD$ with mean luminosity. Lower: $L_X$ vs. time for $\AEp$ %with mean luminosity. Data from Table \ref{4472fits}.} 
%\label{fig:D_E}
%\end{figure}

\subsubsection{Short-term (inter-observational) variability}

\label{appendixB}
Two GC ULXs \citep{2007Natur.445..183M,2010ApJ...721..323S} show variability on short time scales. We extracted light curves of GC ULXs from our sample in any observations that had 500 source counts or greater (see Tables \ref{obs4472}, \ref{obs4649}, and \ref{obs1399}). The source $\macc$ shows interesting behaviour in ObsIDs 12888 and 12889. These observations were taken a week apart, and were each near 150ks in length. The fluxes in each observation were significantly different (see Table \ref{4472fits}), and yet within the observation, no clear variability was observed.

\begin{figure}

\includegraphics[scale=0.45]{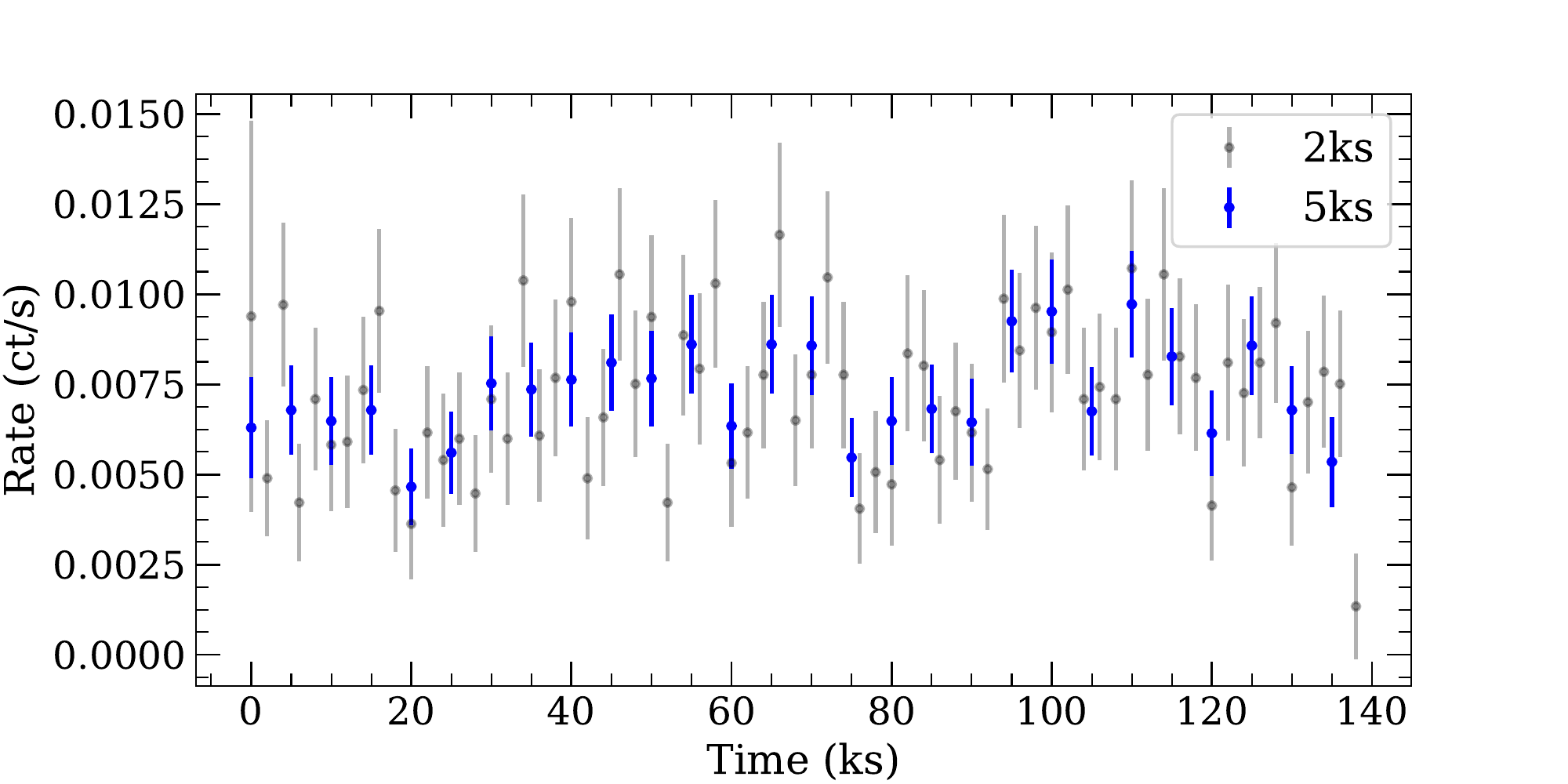}

\includegraphics[scale=0.45]{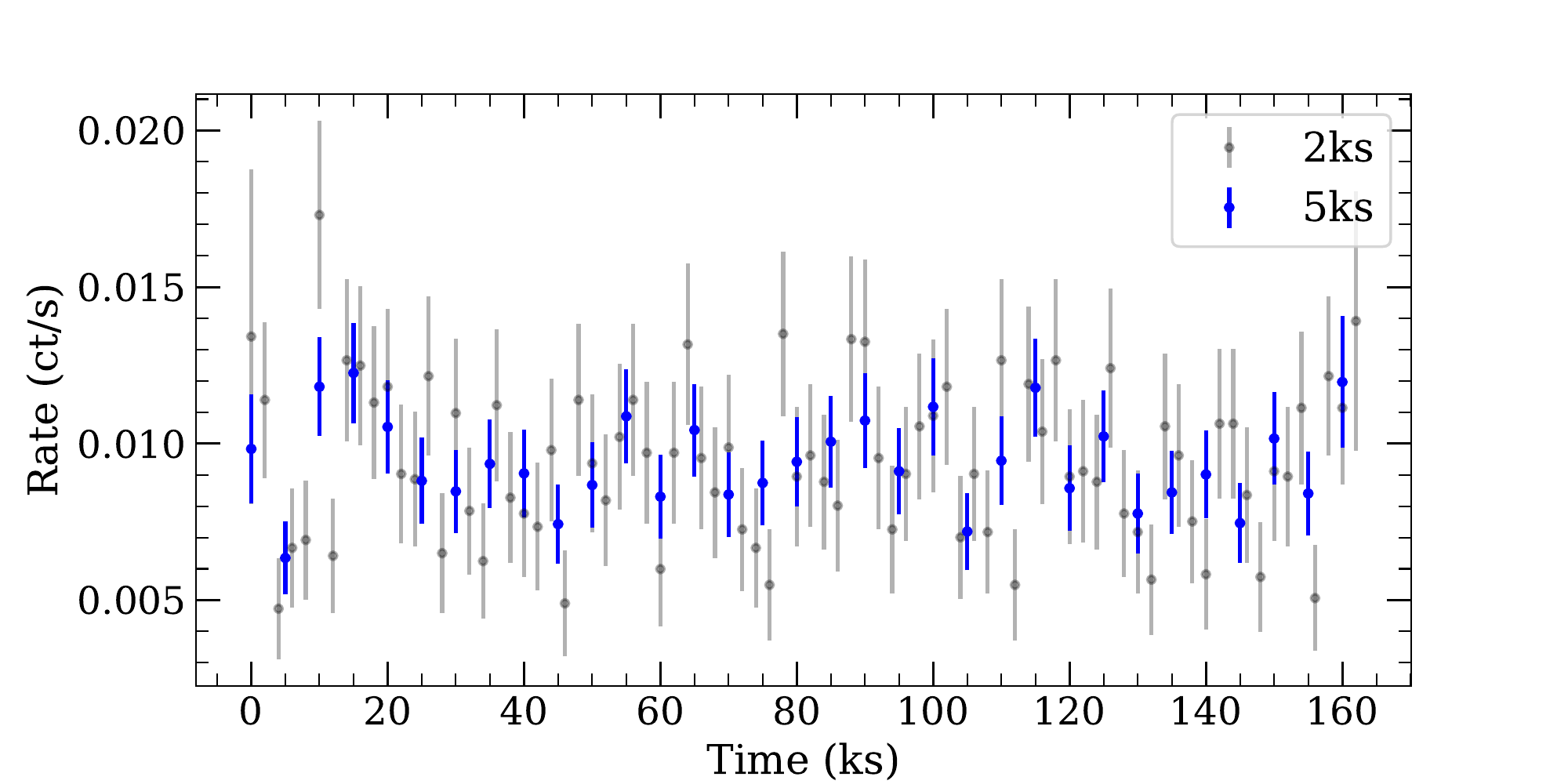}

\includegraphics[scale=0.45]{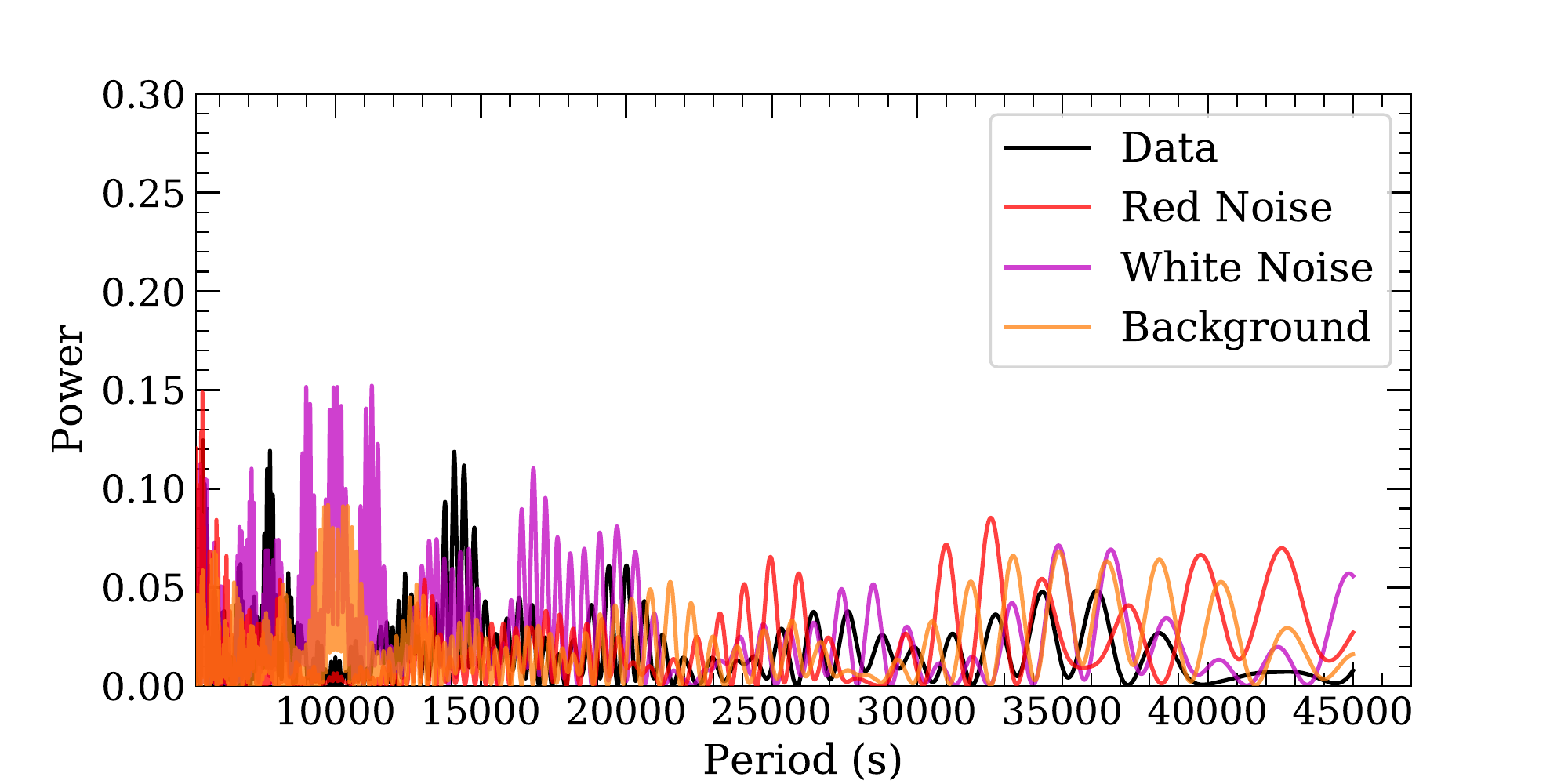}

\caption{Upper: Light curve for $\macc$ from  ObsID 12889. Middle: Light curve for $\macc$ from  ObsID 12888. Both are binned by 5~ksec and 2~ksec. Lower: Lomb-Scargle periodogram of ObsIDs 12888 \&12889 of $\macc$ binned by 5~ksec compared to the Lomb-Scargle  periodograms of the background, as well as red noise and white noise. } 
\label{Fig:322_321}
\end{figure}
We also search for inter-observational variability or periodicities in any \textit{Chandra} light curves with greater than 500 source counts (See Tables \ref{obs4472}, \ref{obs4649}, \ref{obs1399}). We implement a generalized Lomb-Scargle periodogram algorithm \footnote{\url{http://www.astroml.org/modules/generated/astroML.time_series.lomb_scargle.html}} \citep{1976Ap&SS..39..447L, 1982ApJ...263..835S}  on combined long observations of the four sources in NGC~4472 and two in NGC~4649 to search for any trends within the observation. The background subtracted light curves as well as the background were extracted from event files filtered to 0.5-8.0 keV using \textsc{CIAO}'s \texttt{dmextract} tool, and binned by 2~ksec and 5~ksec. 

To determine how significant (if at all) any periods identified in the periodogram are, we computed the periodograms of red noise, white noise and the background and compare them to the periodogram of the data for bins of 2~ksec and 5~ksec. We use the \textsc{DELCgen} package \citep{2016ascl.soft02012C} to generate red noise simulations of the \textit{Chandra} light curves \citep{1995A&A...300..707T}. 
We also extract background light curves and compute the Lomb-Scargle periodogram in the same manner. For white noise, we  shuffle\footnote{\url{https://docs.scipy.org/doc/numpy-1.15.0/reference/generated/numpy.random.shuffle.html}} our original light curves and re-compute the periodogram. 

We find that there are no clear significant periods in this data and that white noise is the main cause of spurious signals in the Lomb-Scargle periodogram (see Figure \ref{Fig:322_321}, (lower panel) for the various contributions of noise to the detected signals in the data).

\section{RESULTS}
\label{sec:results}

We fit a sample of eight GC ULXs located in NGC~1399, NGC~4472 and NGC~4649 (see Table \ref{all_srcs}) with two different single component models, an absorbed disk or an absorbed power law, and take data from the literature for a ninth \citep{2018arXiv180601848D}. We also consider a power law with intrinsic absorption, but find that for almost all the sources, the intrinsic absorption component was either consistent with zero, or not a statistical improvement over a disk model.

We also consider a two component model, however, it was not enough of a statistical improvement over the single component model. In the few cases where the two component model was a statistical improvement, the power law index was always unphysical.  The $\chi^2$ statistics for the single component models either indicated that one model was a better fit than the other, or the statistics were comparable, except in the case of GCU6, which was ambiguous.

\begin{figure} 
\includegraphics[width=9cm]{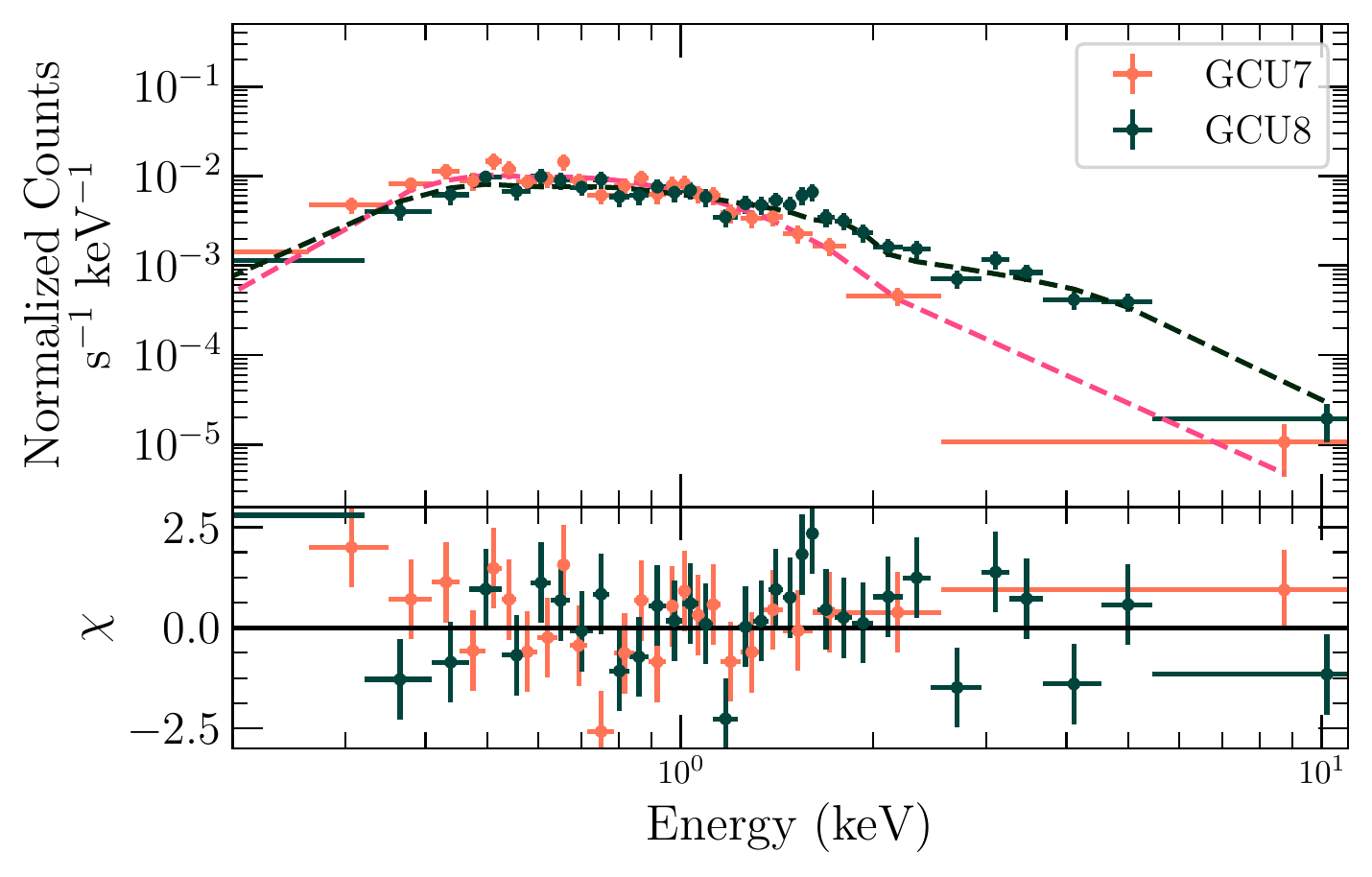}
\caption{Spectra and fitted models (fit residuals on lower panel) of GCU7 and GCU8. GCU7 is best fit by a disk model, while GCU8 is much harder and better fit by a power law model.  } 
\label{fig:spectra}
\end{figure}

Figure \ref{fig:spectra} shows the best fit models and residuals for spectra from GCU7 and GCU8. GCU8 was consistently brighter than GCU7 (see Table \ref{obs1399}), and the best fit model was a power law with no intrinsic absorption. The lower luminosity source, GCU7, was best fit by a \texttt{diskbb} model. In this case specifically, the system producing the emission in GCU8 is physically different than that of GCU7. 

The data never require that the sources change between a disk to a power law between observations, although it is difficult to [strongly] rule out the possibility. The ninth GC ULX source, RZ2109 is the only source that is clearly a two component model \citep{2008MNRAS.386.2075S}; however its power law indices were not well constrained \citep{2018arXiv180601848D}. Like these other GC ULX sources, RZ2109 also shows no strong evidence for changes in spectral state. 

\begin{figure*}
\includegraphics[width=14cm]{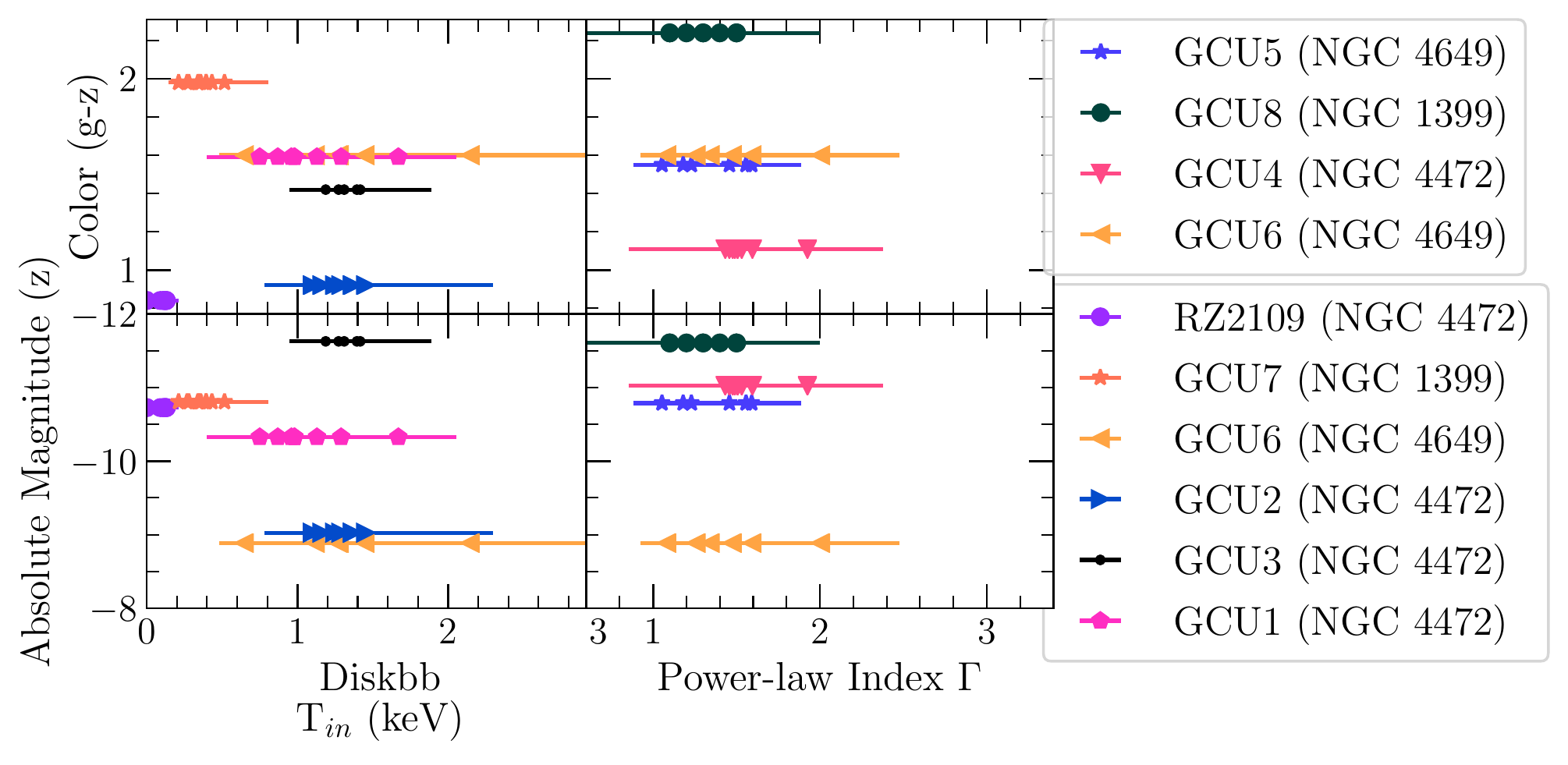}

\caption{Upper: Plot of optical cluster colour ($g-z$)  versus best fit spectral parameter for GC ULXs.  Lower: Plot of absolute $z$-band magnitude versus best fit spectral parameter for GC ULXs.  Neither absolute magnitude nor colour appear to be correlated with best fit spectral parameter in this sample. However, it is worth noting how red in colour GCU8 is. (Optical cluster values from Table \ref{all_srcs})} 
\label{fig:color}
\end{figure*}

We also compare optical cluster colour (g-z) and magnitude (z) to the best fit spectral index in Figure \ref{fig:color}. There does not appear to be clear correlation between optical colour and X-ray behaviour of the sources.
%It shows a strong variability in luminosity without clear corresponding changes in disk temperature . 

\begin{figure*}
\includegraphics[width=16cm]{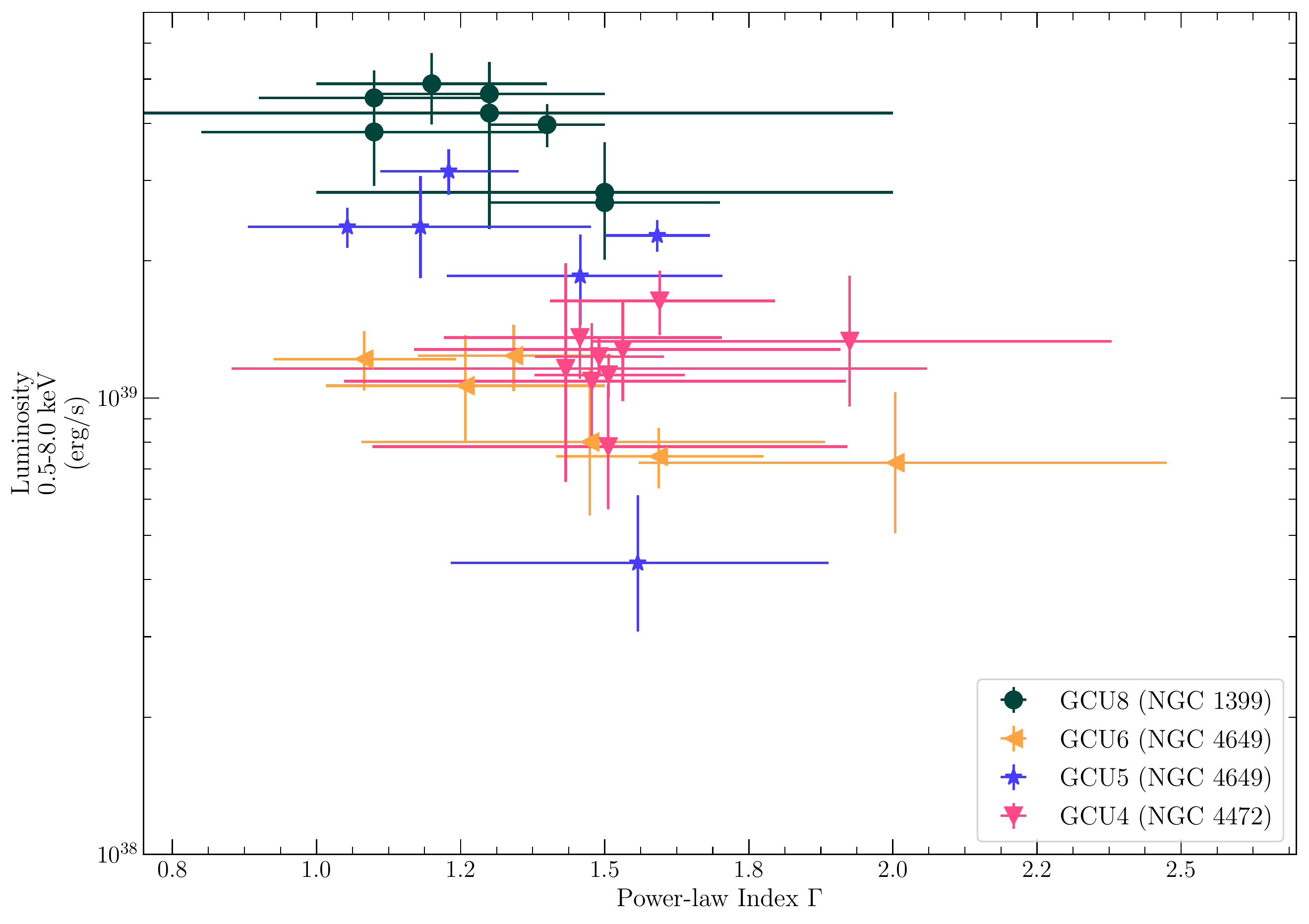}
  \caption{ $\Gamma$ vs. Luminosity (0.5-8.0 keV) for GC ULXs best fit by \texttt{tbabs*pegpwrlw}.  }

  \label{fig:glum}
\end{figure*}

\begin{figure*}
\includegraphics[width=16cm]{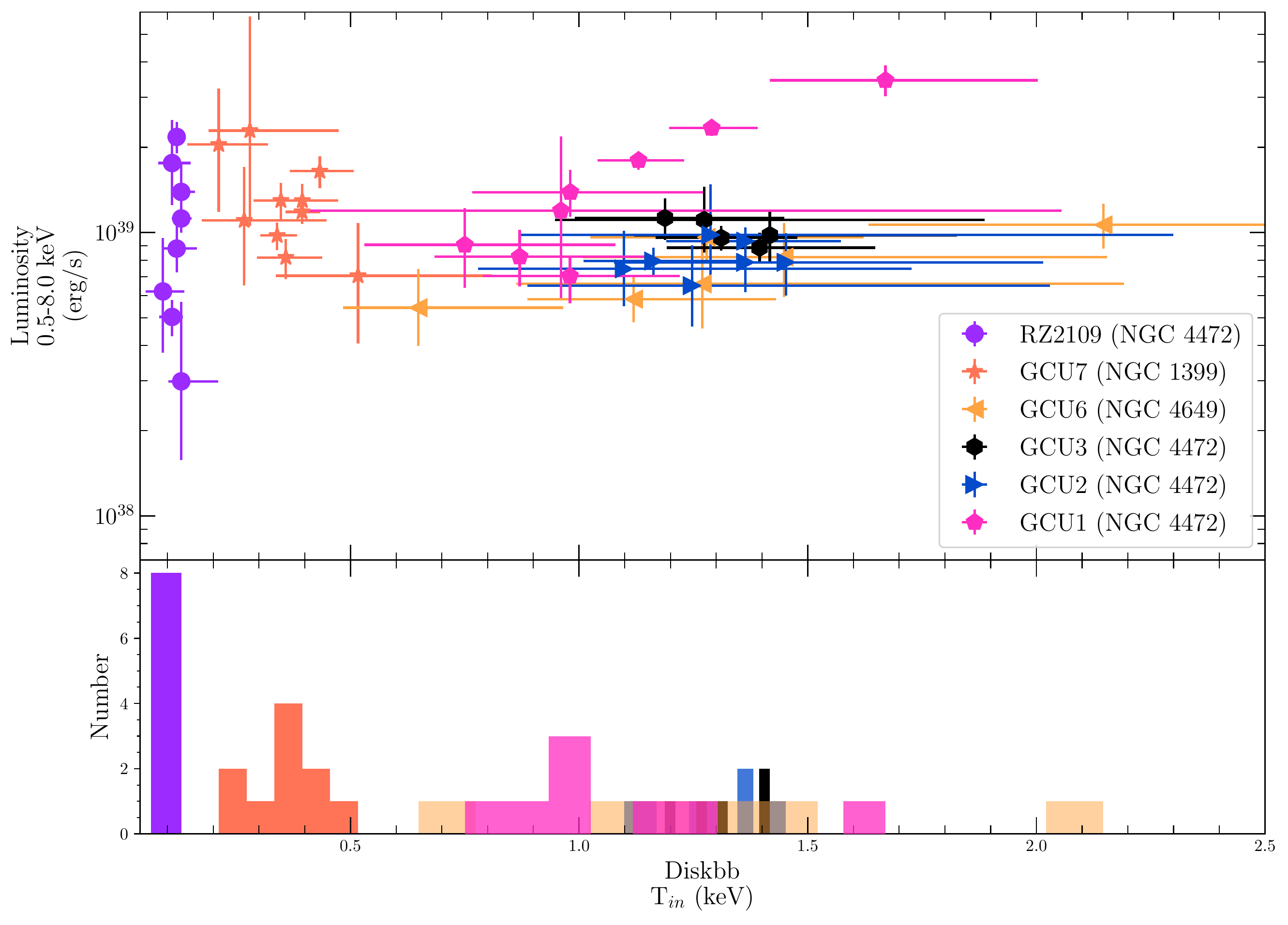}
  \caption{Upper:$T_{in}$ vs. Luminosity (0.5-8.0~keV)for GC ULXs best fit by \texttt{tbabs*diskbb}. Data for RZ2109 is taken from \citet{2018arXiv180601848D}. Lower: Histogram of best fit kT for all sources.}

  \label{fig:tinlum}
\end{figure*}

The ultimate aim of this paper is to study the behaviour of the brightest ($L_X$ $>$ $10^{39}$ erg s$^{-1}$) GC ULXs. To quantify how the spectral parameters of these sources change with their luminosity, we plot the values of the spectral component versus luminosity in Figures \ref{fig:glum} and \ref{fig:tinlum}. Any GC ULX with largely better statistics for a power law was plotted in Figure \ref{fig:glum}; any source with consistently better statistics for a disk is plotted in Figure \ref{fig:tinlum}. $\CC$ was ambiguous as to whether a disk or a power law was a better fit, so we therefore present it on both plots. 

The power law sources show little variability in either $\Gamma$ or luminosity.  Within uncertainties, both $\BC$ and $\AEp$ have similar luminosity and power law index values across all of their own observations, with $\AEp$ at a lower luminosity than $\BC$. $\CC$ also does not vary significantly in either parameter. $\RS$ seems to follow the same behaviour, except for ObsID 785 (2000-04-20), the first observation taken of this source. It seems as though it begins with a low luminosity, then brightens and stays fairly consistent at that luminosity/$\Gamma$.

The disk sources appear to be bimodal: they are either sources with $T_{in}$ greater than 0.5 keV, or much less than that temperature. Of the sources with $T_{in} >$ 0.5 keV, $\macc$ shows the most variability. $\CC$ shows some variability in $T_{in}$  and $L_X$. $\AC$ and $\AD$ do not show significant variability in either luminosity or disk temperature. 

Finally, the two sources with nearly steady disk temperatures below 0.5 keV, RZ2109 at $\simeq$ 0.15 keV and $\irwin$ at $\simeq$ 0.4 keV do show significant variability in luminosity while having no large change in $T_{in}$. It is of note that the only sources that vary significantly with $L_X$ but not visibly with the spectral parameter are RZ2109 and $\irwin$. Interestingly, both RZ2109 and $\irwin$ show optical emission lines \citep{zepf08,irwin2010}.

To determine the extent to which variations in luminosity and variations in spectral fit parameter (either kT or $\Gamma$) are correlated and estimate the correlation slope, we determine the best fit line to the data for each individual source. To carry out this fit, we use \textsc{linmix}\footnote{Python port by J. Meyers: \url{https://github.com/jmeyers314/linmix}} \citep{linmix} which uses Bayesian inferences, and develops MCMC sampling to allow linear fits while accounting for uncertainties in both variables. However, this implementation does not allow asymmetric parameter uncertainties which typically rise in X-ray spectral fits. Thus, we conservatively chose the larger uncertainty value on each parameter for both lower and upper values. We fit the correlation with $T_{in}$ as a function of $L_X$, which allows a simple test for lack of correlation for systems which luminosity seems to vary independent of disk temperature (RZ2109 and GCU7). 
\begin{table}
\centering
\caption{Best fit slopes and uncertainties of sources that varied in kT or $L_X$.  }
\label{slopes}
\begin{tabular}{|l|l|}
\hline
\hline
Object & Slope            \\ \hline
RZ2109 & 0.0 $\pm$ 0.02   \\ 
GCU1   & $0.29 \pm 0.16$ \\ 
GCU7   & $0.08 \pm 0.2$   \\  \hline
\end{tabular}
\end{table}

\begin{figure*}
$\begin{array}{rl}
    \includegraphics[width=0.5\textwidth]{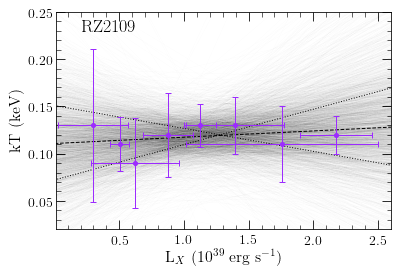}&
    \includegraphics[width=0.5\textwidth]{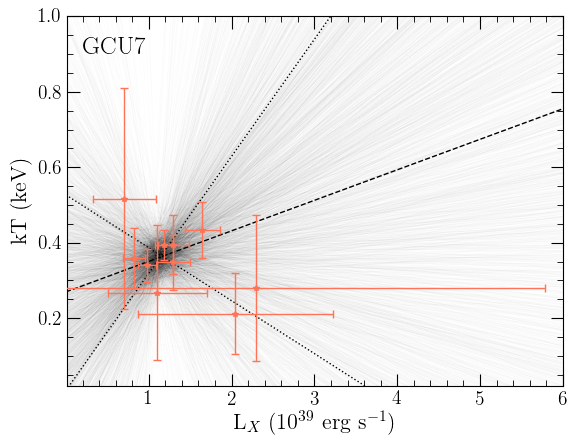}\\
    \multicolumn{2}{c}{\includegraphics[width=0.5\textwidth]{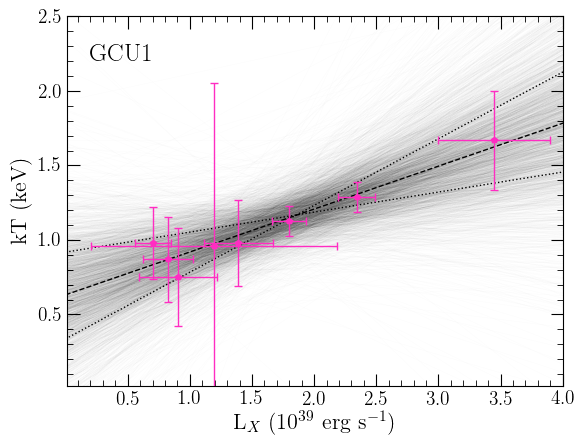}}
\end{array}$
\caption[ ]{\label{fig:linmix} \textsc{linmix} best fits of $L_X$ vs kT for RZ2109, GCU7 (Slopes consistent with zero) and GCU1 (slope inconsistent with zero). }
\end{figure*}

We used this method to fit the slopes of RZ2109, GCU1, and GCU7. The best fit slopes and uncertainties are reported in Table \ref{slopes}. We did not fit slopes for GCU2, GCU3, or GCU6 as they do not appear to vary significantly in either kT or $L_X$. The difference in slopes between the sources below 0.5 keV (RZ2109 and GCU7) and the sources above 0.5 keV (GCU1) is suggestive of a dichotomy between low kT and high kT sources, with the low kT sources having a slope that is likely consistent with zero, and some high kT sources having a non-zero slope.  See Figure \ref{fig:linmix} for the best fit slopes and errors of RZ2109, GCU1 and GCU7.

\section{CONCLUSIONS}
\label{sec:conclusions}
We consider a total sample of nine ultraluminous X-ray sources ($L_X \geq 10^{39}$ erg s$^{-1}$) physically associated with globular clusters for which we do new data analysis for eight and rely on our previous study of RZ2109 for the ninth.  We find that the sources are best fit by a single component -  either an absorbed disk or an absorbed power law. Two component absorbed disk plus power law model is not statistically required for any of these sources. The two component fits also give unphysical power law indices.

When we compare the luminosity to the spectral parameters of the sources, we find that sources best fit as power laws have either no clear variability in either power law index or luminosity, or only show mild variability in each parameter. When comparing luminosity to inner disk temperature, the sources split into two temperature ranges, one group with temperatures above 0.5 keV, and the other with low $T_{in}$. The part of the sample best fit by disks  with temperatures below 0.5 keV and show strong variability in luminosity with no clear variability in disk temperature. The disk sources with temperatures below 0.5 keV are the only ones to show variability only in luminosity, without significant corresponding spectral changes. 

Prior work \citep{2001ApJ...557L..35A, 2002ApJ...574L...5K,2003ApJ...595..743S,  2004ApJ...613..279J} has shown that more luminous (and presumably more massive) globular clusters are more likely to contain low mass X-ray binaries. Beyond that, our GC ULX sample does not show any clear correspondence of X-ray behaviour or luminosity to the optical photometric properties of the cluster. However, some  aspects of the X-ray behaviour appear to correlate with the presence of optical emission lines. RZ2109, $\irwin$ and $\RS$ all have published optical spectra. RZ2109 shows bright and broad [OIII] emission lines beyond the globular cluster continuum, with no hydrogen emission detected \citep{zepf08}. $\irwin$ has narrow [OIII] and [NII] emission lines beyond the cluster continuum, also with no hydrogen emission \citep{irwin2010}. RZ2109 and $\irwin$ both have similar behaviours in X-ray, with consistent, low disk temperatures and luminosity variability. 

$\RS$ shows no optical emission lines at all, including no hydrogen emission lines. $\RS$ has very different X-ray behaviour, as it has a better fit as a single power law model that does not show the same kind of luminosity variability seen in the other two sources. Additionally, when it is fit as a single disk component, it has temperatures greater than 1 keV \citep{2012ApJ...760..135R}. We note that none of the three sources has hydrogen emission present.

We postulate that X-ray behaviour may be linked with optical emission, as the sources with low disk temperatures both have optical emission, but the source with no optical emission has a vastly different behaviour in X-ray. However, optical follow-up on other sources in our sample is necessary to confirm such a claim. 

\section{Acknowledgments}

KCD, SEZ, and MBP acknowledge support from Chandra grant GO4-15089A  and NASA grant number HST-AR-13923.001-A from the Space Telescope Science Institute, which is operated by AURA, Inc., under NASA contract NAS 5-26555.
SEZ and MBP also acknowledge support from the NASA ADAP grant NNX15AI71G.
This research has made use of data obtained from the Chandra Data Archive and the Chandra Source Catalog.  We  also acknowledge use of NASA's Astrophysics Data System and Arxiv. The authors thank Claire Kopenhafer, Jay Strader and Ryan Urquhart for helpful discussion. We also thank the anonymous referee for their suggestions to improve the paper.

The following softwares and packages were used for analysis: \textsc{ciao}, software provided by the Chandra X-ray Center (CXC),   \textsc{heasoft} obtained from the High Energy Astrophysics Science Archive Research Center (HEASARC), a service of the Astrophysics Science Division at NASA/GSFC and of the Smithsonian Astrophysical Observatory's High Energy Astrophysics Division, SAOImage DS9, developed by Smithsonian Astrophysical Observatory, \textsc{astroML} \citep{astroML}, \textsc{linmix} \citep{linmix}, \textsc{numpy} \citep{2011arXiv1102.1523V}, \textsc{DELCgen} \citep{2016ascl.soft02012C},  \textsc{matplotlib} \citep{2007CSE.....9...90H}, Palettable\footnote{\url{https://jiffyclub.github.io/palettable/}}, \textsc{scipy} \citep{scipy}, and \textsc{astropy} \citep{2013A&A...558A..33A}. 

%\clearpage
%\appendix
%\section{F-test Values for GC ULXs}
%\label{appendixstat}
%Figures \ref{4472_ftest}, \ref{4649_ftest} and \ref{1399_ftest} show the F-test probability values when comparing either single component model, \texttt{tbabs*diskbb} or \texttt{tbabs*pegpwrlw}, to the two component model, \texttt{tbabs*(diskbb+pegpwrlw)}. We compared both single component models to the two component model for each source in observations with greater than 100 source counts. Section \ref{sec:observations} discusses how we used these results to determine that the data are each better fit by single component models rather than two component.

\bibliographystyle{mnras}

\bibliography{ulxbh}

% Don't change these lines
\bsp	% typesetting comment
\label{lastpage}
\end{document}